\documentclass[11pt,final]{article}
\usepackage{amsfonts}
\usepackage{amsmath}
\usepackage{amssymb}
\usepackage{amsthm}
\usepackage[margin=1.0in]{geometry}
\usepackage{graphicx}
\usepackage{float}

\newcommand{\bE}{\mathbb{E}}

\newcommand{\bH}{\mathbf{H}}

\newcommand{\freq}{\bH}

\DeclareMathOperator*{\argmin}{arg\,min}

\begin{document}

\title{Fitness Estimation for Genetic Evolution of Bacterial Populations}
\author{Sergey S. Sarkisov\footnote{University of Houston, Department of Mathematics, Houston, TX 77204-3008}, 
Ilya Timofeyev\footnote{Corresponding Author; University of Houston, Department of Mathematics, Houston, TX 77204-3008; email: ilya@math.uh.edu}, 
Robert Azencott\footnote{University of Houston, Department of Mathematics, Houston, TX 77204-3008; email: razencot@math.uh.edu}}
\maketitle

\begin{abstract}
In this paper we develop and test algorithmic techniques to estimate genotypes  fitnesses by analysis of observed daily frequency data monitoring the long-term evolution of bacterial populations. In particular, we develop a non-linear least squares approach to estimate selective advantages of emerging new mutant strains in 
locked-box stochastic models describing bacterial genetic evolution 
similar to the celebrated Lenski experiment 
on Escherichia Coli.
Our algorithm first analyses emergence of new mutant strains
for each individual trajectory. 
For each trajectory our analysis is progressive in time, and successively focuses on the first mutation event before analyzing the second mutation event. 
The basic  principle applied here is to minimize (for each trajectory) the mean squared errors of prediction $w(t) - W(t)$ where the observed white cell frequencies $w(t)$ are predicted by $W(t)$, which is computed as the conditional expectation of $w(t)$ given the available information at time $(t-1)$. The pooling of all selective advantages estimates across all trajectories provides histograms on which we perform a precise  peak analysis to compute final estimates of selective advantages. We validate our approach using ensembles of simulated trajectories.
\end{abstract}

\noindent
\hspace*{0.95cm}

\noindent
\hspace*{0.95cm}


\setcounter{page}{1}

\section{Introduction}
\label{sec:intro}
Studies of evolutionary dynamics, including bacterial populations, received a considerable amount of attention in recent decades starting with the celebrated Lenski experiment \cite{Lenski1991,len94a}.  Since then other groups contributed to experimental studies of  Escherichia Coli (see e.g. \cite{Cooper2001, Hegreness2006, Barrick2010, Peng2017, Deatherage2017, Illinworth2012, Gordo2013}). 
Typical setups for these experiments involve multiple populations processed in parallel. These populations undergo daily growth followed by daily selections of fixed-size samples. Initially, all populations consist only of cells having the same "ancestor" genotype. Since most evolutionary experiments are carried out for a long time, a varied range of random mutations occur in different populations, so that the genetic composition of different populations can differ drastically over time. The main goal of such experiments is to understand major features of evolutionary dynamics, including the rate of evolutionary change, likelihood of emergence of a new genoype, fitness landscape, probabilities of fixation for emerging mutants, etc.

The  Lenski  long-term evolution experiments on Escherichia coli  provided a considerable insight and evidence  for the mechanisms of growth and adaptation of asexual organisms  (e.g. \cite{Lenski1991,len94a,Fox2015}).  Analysis of the experimental data shed some light upon  the relative growth rates of stronger mutants and their ancestors. The  Lenski experimenst primarily focuses on the overall population adaptation \cite{Barrick2013,Woods2011}. Our goal is to develop and test an efficient algorithm dedicated to estimating individual selective advantages for emerging mutants.

The adaptive evolution of bacterial populations is driven by a random emergence of beneficial mutations and spread of the fittest mutants due to natural random selection.  Various mathematical models  of this evolution process have been developed (see e.g. \cite{Kimura1962, Simon2013}), and estimation of process parameters in these models is always one of the key issues.
The evolutionary dynamics of bacterial populations depends on two family of parameters, namely the  selective advantages of mutants and the rate of mutations \cite{Hegreness2006,Perfeito2007,Lang2011}.  
In this paper we assume that the mutation rate $\mu$ is roughly the same for all emerging mutants, and is already known. Therefore, we focus on estimating selective advantages for multiple competing mutants. Our techniques can be easily extended to estimate $\mu$ as well. Several previous papers addressed similar questions. For instance, one study involved both experimental and theoretical analyses of evolving viral populations \cite{Rokyta2005,Rokyta2008}. These  papers include estimation for their adaptation models parameters, as well as for the distribution of beneficial mutation effects. However, this work was limited to viruses, because of their small genome size, high mutation rate, and large mutation effects.   By contrast, in bacterial populations, direct estimates of mutational parameters is much more complex \cite{Rozen2002} because these parameters are usually estimated indirectly by inference from observable markers \cite{Rozen2002,Hegreness2006,Barrick2010}.
A more recent study, involving  Tim Cooper's experiments on Escherichia Coli, focused on populations where a single mutant type emerges and overtakes the population \cite{Zhang2012}.  Mathematical derivations and intensive model simulations were combined to derive accurate estimates for the mutation rate and selective advantage of the first mutant reaching fixation. Another earlier parameter estimation study \cite{Hegreness2006} used non-linear regression to analyze the first emerging mutant. Our approach here is complementary to these earlier studies. In particular, we develop efficient parameter estimation algorithms to cover the emergence and competition of multiple mutant types.

In this paper, we consider mathematical setup motivated by the Escherichia Coli  evolutionary experiments \cite{Zhang2012, Hegreness2006, Barrick2010}. In such experiments one observes many populations simultaneously evolving in parallel. Each initial population consists entirely of $N$ cells with the same "ancestor" genotype, with $N$ typically ranging from 50,000 to 100,000. In each population, half of the cells are colored via a particular marker (e.g. Ara$^+$) and the other half are colored via a different marker (e.g. Ara$^-$). These two markers define "white" and "red" cell colonies, initially coexisting in proportions 50/50 in all populations. Cell dynamics is not altered by these markers.  Asexual reproduction occurs  via binary fission, and  genetic markers are passed down from parents to offspring \cite{Barrick2010}. The populations are allowed to grow daily until exhaustion of the daily initial glucose or equivalent "food supply", which happens in less than 24 hours. At that point each cell population reaches a size in the range of $10^7$ to $10^8$, and a sub-population of approximate size $N$ is extracted by dilution and transferred to a new culture of fresh growth medium.  This transfer step is repeated daily for all populations. The white and red cell frequencies are estimated periodically for instance by plating cells on indicator media.
Therefore, 
after initialization of all populations with identical "ancestor" genotype and  50/50 marker coloration into "white" and "red" colonies, the daily cycle of  growth, mutations, and dilution is repeated for hundreds of days.  A beneficial mutation emerging  on day $t$ among  white cells can be detected at a later time $t+s$ as soon as it generates a significant upward change in the frequency $w(t+s)$ of white cells. Similarly beneficial mutations emerging among red cells at time $t$ can be detected by significant downward change in $w(t+s)$. 
Ideally the whole trajectory $w(t)$ can be recorded for each population. After a rough detection for the emergence of new mutant strains, we apply a non-linear least squares method to estimate more accurately the times of emergence of these new strains, as well as their selective advantages. This is done separately for each trajectory. We then combine
estimated selective advantages computed for each individual trajectory. 
By a detailed analysis of the histogram of all these estimates, we finalize more accurate estimates for the selective advantages of the observed mutant strains.
Our stochastic model for the evolutionary dynamics of bacterial populations is a  Markov chain where the cycle of growth, mutation, selection is repeated daily. To implement our non-linear least squares estimation of model's parameters, we develop explicitly computable predictors of next day white cell frequency, and minimize the mean squared errors of these predictors with respect to the observed data. 
We validate our approach on simulated data. In particular,  we use Markov chain model to numerically generate an ensemble of evolutionary trajectories. We then apply our estimation algorithm to these simulated trajectories and evaluate accuracy of our estimated  selective advantages. We utilize Gaussian mixture approach to analyze numerical results, which validate the statistical consistency of our parameters estimators.

\section{Stochastic Model}
\label{sec:stochmodel}

Various modifications of Markov chain ``locked-box'' models have been often used to describe
evolutionary experiments outlined above 
(e.g. \cite{Zhang2012,Rice2004,Hegreness2006,Ross1996,Simon2013}). 
In these models a daily random mutation phase occurs right after a daily deterministic growth phase, and is followed by a random selection phase, formalized by extraction of a random sample of fixed size $N$. The evolutionary  experiment is then modeled by repeating the daily cycle
 \textit{growth} $\Rightarrow$ \textit{mutations} $\Rightarrow$ \textit{selection} $\Rightarrow \ldots$.
 
We consider only a finite number $i \in \{1,2,3, \ldots , g\}$ of possible genotypes. The ancestor genotype is denoted as $i=1$ and the other $(g-1)$ genotypes are mutants. Let $t \in \{1,2, \ldots \}$ be the discrete time index (measured in days). Denote  $h_i(t)$, the frequency of genotype $i$ at the beginning of day $t$. The random vector $h(t) = \{h_1(t), \ldots, h_g(t)\}$ clearly verifies $\sum_{i=1}^{g}h_i(t) = 1$, and is the whole histogram of genotypes frequencies at the beginning of day $t$.  Next, we describe the three daily phases in more detail. First, we describe the Markov chain model
for the evolution of a single-color population; extension to two colors is done in a rather straightforward manner by doubling the number of cell types.

\subsection{Daily Growth Phase}
At the beginning of any day $t$, the current cell population always has total size $N$, and is partitioned into $g$ sub-populations $POP_i(t)$ consisting of cells with the same genotype $i$. Let $N_i(t) = N \times h_i(t)$ be the size of $POP_i(t)$, so that $N = \sum_{i=1}^{g} N_i(t)$. The daily growth phase is modeled by a deterministic exponential growth in which cell divisions occur at fixed exponential rate $\rho_i$ for all cells of genotype $i$. Therefore, 
at the end of the day $t$ growth phase, which has fixed duration $D$, the size of $POP_i(t)$ becomes 
$$
S_i(t) = F_i \times N_i(t) = N F_i h_i(t), 
$$
where the multiplicative daily growth factor $F_i = \exp(\rho_i D)$ is specific to genotype  $i$.  Hence, the total population size at the end of the day $t$ growth phase is  $S(t) = \sum_{i=1}^{g} S_i(t)$.
Without loss of generality, assume that genotypes are ordered with respect to their growth factors so that $F_1 < F_2 < \dots < F_{g-1} < F_g$. The \textit{selective advantage} $s_i$ of genotype $i$ with respect to the ancestor genotype $1$, is defined by
\begin{equation} \label{sj} 
F_i = F_1^{(1+s_i)} 
\qquad
\text{or equivalently} 
\qquad
s_i = \frac{\log(F_i)}{\log(F_1)} - 1.
\end{equation} 
Some models (see e.g. \cite{Desai2007, Zhang2012}) assume that the restricted daily amount of nutrients forces the  daily population growth to stop when population reaches  a fixed large maximum size.  However, in this paper, our model assumes that the daily growth phase has a fixed duration $D$, already integrated in the multiplicative daily growth factors $F_1, \ldots, F_g$. At the end of the day $t$ growth phase one has $N F_1 \leq S(t) \leq N F_g$. 
%

\subsection{Daily Mutations Phase}
At the end of  the day $t$ growth phase, the population size $S(t)$ is typically several orders of magnitude larger than the daily initial population size, $N$. 
For instance,
in Escherichia Coli long term experiments $N$ ranges from $10^{5}$ to $10^{8}$, and $F_1 \approx 200$, while the mutation rate $\mu$ ranges from $10^{-7}$ to $10^{-6}$. Therefore, $S(t)$ ranges from $10^{7}$ to $10^{10}$, and every day a positive number of new mutants can be expected to emerge during the daily mutation phase. Here we consider only beneficial mutations, since deleterious mutations are typically quickly eliminated  from the population due to a lower selective advantage.

In our mathematical model, we assume that the daily mutations are independent random events, simultaneously occurring at the end of the daily growth phase. Next, we describe precise mathematical formulation for mutations.

At the end of  the  day $t$ growth phase, and just before the day $t$ mutations phase, the day $t$ initial population $POP_i(t)$  of cells with genotype $i$, has grown into  a much larger population $GPOP_i(t)$, of size $S_i(t)$. Denote $\nu_{i}= \nu_i(t)$ the random number of mutants emerging from population $GPOP_i$ during the mutations phase. We assume that the conditional distribution of $\nu_i(t)$ given the initial day $t$ histogram $h(t)$ is a  Poisson distribution with mean
\begin{equation}
\label{meanmut}
    \lambda_{i} =\bE \left[ \nu_i (t) | h(t) \right]= \mu S_i(t) = \mu N F_i h_i(t).
\end{equation}
Note that the mutation rate $\mu$ is assumed to be the same for all genotypes. Our approach can be easily extended to mutation rates  $\mu_i$ which depend on genotype $i$, provided all $\mu_i$ are small and are of the same order of magnitude.
The conditional distribution of $\nu_i$ is hence given by 
\begin{equation}
\label{Pjmut}
Pr(\nu_i=k) = e^{-\lambda_i} \frac{\lambda_i^{k} }{k !}.
\end{equation}
We assume that all $\nu_i(t)$, with $1\leq i \leq g$ are conditionally independent given the histogram $h(t)$.

Our mutation model also involves a fixed  (unknown) transition matrix $\mathcal{P} = \{p_{i,j}\}$ with $0\leq p_{i,j}\leq 1$ and $\sum_{j=1\ldots g} p_{i,j} = 1$. The matrix  $\mathcal{P}$  governs  the stochastic transition from genotype $i$ to genotype $j$ by a single random mutation. Hence  for each $i$, any mutant emerged from $GPOP_i(t)$ has probability $p_{i,j}$ of having genotype $j$. 
Of course one imposes $p_{i i}= 0$ for all $i$.

Denote $\nu_{i,j} = \nu_{i,j}(t)$ the random number of mutants emerging from population $GPOP_i(t)$ and mutating into cells of genotype $j$. Then we have
$$
\nu_i = \sum\limits_{j=1}^g \nu_{i,j} 
\quad \text{and}  
\quad \nu_{i,i}= 0.
$$
The conditional distribution of the random vector $\left[\nu_{i,1}, \ldots , \nu_{i,g} \right]$ given both  $h(t)$ and $\nu_i(t)$  is then naturally assumed to  be a \emph{multinomial distribution} with probabilities $\{p_{i,1}, \ldots p_{i,g}\}$ within a population of size $\nu_i(t)$. Hence for each $i$ and for all non-negative integers $ k_1, \ldots, k_g$ verifying $k_i= 0$ and $ \sum_{j = 1 \ldots g} k_j = \nu_i$, we have 
\begin{equation}\label{mut}
\begin{gathered}
Prob(\{\nu_{i,1}= k_1, \ldots, \nu_{i,g}=k_g \}\; | \; \{\nu_i, h_i\}) = 
\nu_i! \prod_{j=1}^g  \frac{p_{i,j}^{k_j}}{k_j!}, \\
\bE[\nu_{i,j} | \{\nu_i(t), h(t)\}] = p_{i,j} \nu_i, \qquad 
\bE[\nu_{i,j} | h(t)] = p_{i,j} \mu N F_i h_i(t).
\end{gathered}
\end{equation}
For any fixed genotype $i$,  denote $EM_i$ and $IM_i$ the respective numbers of "emigrants" from population $GPOP_i$ and of "immigrants" into $GPOP_i$. More precisely let 
\[
EM_i =  \sum_{j = 1}^{g} \nu_{i,j} \quad \text{and} \quad  
IM_i =  \sum_{j = 1}^{g}\nu_{j,i}.
\]
Let $\Delta S_i(t) = IM_i - EM_i$ be the net change in the size of population $GPOP_i$. One can easily check that the mutation phase does not alter the total population size, i.e.
$\Delta S(t) = \sum_{i=1}^g \Delta S_i(t)  = 0$.
%

\subsection{Daily Selection Phase}
On each day $t$, after the growth phase and the mutations phase, a \emph{random selection} is performed by extracting a random sample of \emph{fixed size} $N$ from the whole terminal population on day $t$, which still has  size $S(t) >> N$. In practical laboratory experiments, this selection is often implemented by culture dilution.

Denote $f_i$ the frequency of genotype $i$ after growth and mutations. We clearly have
\[
f_i = f_i(t) = \frac{N F_i h_i(t)  + \Delta S_i(t)}{S(t)}.
\]
The daily  random selection of $N$ cells  within a day $t$ terminal population of size $S(t)$ is then modelled by a multinomial distribution based on the probability vector $f_1, \ldots, f_g$.
In particular, 
let $\kappa_i$ be the number of genotype-$i$ cells after the day $t$ random selection. The joint distribution of $\kappa_i$, $i=1,\ldots,g$ is given by
\begin{equation}  \label{dilut}
Prob \big(\kappa_1 = n_1, \ldots , \kappa_{g} = n_g \big) =  
\left( \frac{N!}{n_1! \cdots n_g!} \right) f_1^{n_1} \cdots f_{g}^{n_g}, 
\end{equation}
where $\sum_{i=1}^{g} n_i = N$.

The $N$ cells randomly selected at the end of day $t$ have a random histogram $h(t+1)$. After physical or virtual "transfer"  these $N$ cells initiate the day $t+1$ cycle of growth / mutations / selection. Thus the day $t+1$ cycle begins with a population of size $N$ and  histogram $h(t+1)$.

The cycle  of \{deterministic growth, random mutations, random selection\} is repeated daily for the whole duration of the long-term  experimental run. The process modeled by these successive three steps is a discrete-time Markov Chain driving  the random vectors $h(t) = (h_1(t), \ldots, h_g(t))$ which describe the time-evolution of genotype frequencies. All frequencies sum up to one, i.e. $\sum_{i=1}^g h_i(t) = 1$, and a sample path of this process can be easily simulated  given the vector of initial genotypes frequencies $h(1)$.

\subsection{Cells Coloring}

A key element in the type of evolutionary experiments we model and study here, is to introduce an artificial color feature inherited through all divisions and mutations without altering major cell properties. In many evolutionary experiments and studies (see e.g. \cite{Plank1979,Korona1994,Hegreness2006,Sousa2013, Zhang2012}) this binary coloring is implemented by pairs of inheritable biomarkers which do not alter cells biology. The initial white and red coloring generates two sub-populations (i.e. white and red) with independent evolutionary dynamics and naturally  doubles the number of frequency variables to $2 g$ in the model, since on each day $t$ we then have $g$ genotype frequencies among white cells and another vector of $g$ genotype frequencies among red cells. 
The model presented above for the single-color evolution  can be immediately extended to the two-colors evolution  by doubling the number of frequencies in the frequency vector $h$.  Therefore, in the two-colors stochastic model, for each day $t$ and each $i \in \{1,2,\dots, g\}$ we denote   $H^w_i(t)$ and $H^r_i(t)$  the respective frequencies of genotype-$i$ in the white and red populations. Define then $\bH^w(t) = \{H^w_1(t),\ldots, H^w_g(t)\}$ and $\bH^r(t) = \{H^r_1(t),\ldots, H^r_g(t)\}$. The state of the whole population is then described  by the $2 g$ dimensional vector of  genotype frequencies $\bH(t)=\{\bH^w(t), \bH^r(t)\}$.

Because cells cannot change color by mutations, daily "growth+mutations" phases evolve independently for the white and red cells, and can hence be analyzed or simulated  separately. But the daily selection step depends on the full "white + red" population after growth and mutations. Thus, formula \eqref{dilut} has to be modified appropriately to describe the random daily selection of $N$ cells according to a multinomial distribution base on  $2g$ frequencies $f_i$, where $i=\{1,\ldots,g\}$ indexes the $g$ white genotypes and $i=\{g+1,\ldots,2g\}$ indexes the $g$ red genotypes.

\section{Parameters Estimation of Stochastic Population dynamics}
\label{sec:method}
\subsection{Formal Evolution Experiments}

The type of genetic evolution experiments we consider here involves $n$ independent populations evolving in parallel. This generates $k$ independent evolutionary trajectories $pop^k(t)$, $k=1,\ldots,n$. Each population at time $t=0$ starts from the same initial condition, namely having  the same size $N$, containing only cells of genotype 1  (ancestor genotype), and colored 50/50 with red and white biomarkers. 

Each population $pop^k(t)$ evolves in  time and successive random emergence of beneficial mutations induce shifts in the relative proportion of  white and red cells. Visibility of these  shifts depends of course on the selective advantages of the emerging mutants. For each stochastic population trajectory $pop^k(t)$, there are the only two observables at each time $t$ - the frequencies of red and white cells.
This is because detailed genotype analysis is expensive and cannot be performed for each $t$. 
For each stochastic trajectory $pop^k(t)$, denote  $\{\tau_1, \tau_2, ...\}$ the days of emergence of new mutants, and  $\{g_1, g_2, ...\}$ as their respective genotypes. Of course, these two sequences of random variables are not directly observable.

Therefore, 
our goal here is to develop accurate estimators for the mutations emergence times  $\{\tau_1, \tau_2, ...\}$ and the selective advantages  $\{s_{g_1}, s_{g_2}, ...\}$ of the emerging mutant genotypes. 
We will focus on trajectories which have only a few mutations before overwhelming fixation of a strongly beneficial mutant. This is motivated by the Tim Cooper experiments on Escherichia Coli (see \cite{Zhang2012}) where  most population trajectories exhibited a mutant fixation after one or two mutations. Trajectories with three or more mutations were fairly rare within 200 days runs, as could be also verified by intensive simulations. Moreover, for trajectories with 4 or more successive mutations, the analysis  of later emerging mutants is a difficult task  due to a high sensitivity of later mutations analysis to estimation errors on the parameters of earlier mutation events.
Our simulations indicate that when a reasonably large number of observed trajectories are available, systematic analysis of the first two or three mutation events can provide fairly accurate estimates of the selective advantages of emerged mutants.

Some mutation events occurring during day $t$ can remain unobservable due to the fact that new mutants can disappear during the random selection at the end of day $t$. 
However, some mutants strains emerging at the beginning of day $t$ may grow fast enough during day $t$ and survive the selection bottleneck at the end of day $t$. Thus, such mutants 
would grow even more on day $t+1$ and affect the overall composition of the population 
(i.e. there will be a shift in the composition of white/red cells).
Denote $\tau_1 <\tau_2 <...$ the successive random times at which a new mutant strain  emerges and persists for at least a few  more days. For each observed population trajectory, we first implement an
approximate estimation  of  times $T_0$, $T_1$, $T_2$, $T_3$, etc. which contain mutation events, i.e. $T_0 < \tau_1 <T_1 <\tau_2 < T_2 < \tau_3 < T_3$, etc. Next, we implement a non-linear least squares estimations of mutation times $\{\tau_1, \tau_2, ...\}$ 
and of the selective advantages during each mutation  $\{s_{g_1}, s_{g_2}, ...\}$.
The main mathematical tool is the derivation of explicit 
analytical expressions for the most likely population histogram $h(t+1)$ given the (observed) 
histogram $h(t)$. 


\section{Estimation of approximate intervals for the first and second mutation}
\label{sec:T1}
First, we estimate approximate intervals for the first and second mutation events. This is carried out for each evolutionary trajectory and these intervals are then used in the optimization algorithm to compute the precise estimates for the 
first and second mutation times. Let $\tau_1$, $\tau_2$ be the emergence times of the first and second mutation events. Then our goal is to find $T_0$ and $T_1$ such that $T_0 < \tau_1 < T_1 < \tau_2$.

Let $w(t)$ and $r(t)= 1- w(t)$ be the frequencies of white and red cells on day $t$, respectively.  Since the white and red cells initially have the same genotype $1$, this genetic composition persists for  $t \leq \tau_1$, so that one must observe $w(t) \approx r(t) \approx 1/2 $ for $t \leq \tau_1$ with fairly small random oscillations of $w(t)$ and $r(t)$ around 1/2. These small oscillations arise due to the random dilution step, since random
selection of $N$ cells after growth can alter slightly the proportion of white and red cells.
For the Escherichia Coli experiments of T. Cooper \cite{Cooper2001} the empirical standard deviation  $\sigma_0$ of these oscillations can be crudely estimated on the first 30 days since the first beneficial detectable mutation events never occur before 30 days.

We define $T_0$ as the last day $t \geq 1$ such that  $|w(u) - 1/2 | < 1.25 \, \sigma_0$ 
for all $1\leq u \leq t$.
With high probability, the first beneficial mutation event occurs at some unknown time $\tau_1 > T_0$. Then for $\tau_1 <t \leq \tau_2$ one of the two frequencies $w(t)$ or $r(t)$ starts increasing at a faster rate than the other (i.e. $| w(t) -1/2| $ increases). 
Hence  we define $T_1$ to be the first day $t$ such that $| w(u) - 1/2 | > 2 \, \sigma_0 $ for $u=t$, $t+1$, $t+2$ and all the $(w(u) - 1/2) $ have the same sign. If $(w(u) - 1/2)>0$ (or $<0$) then we conclude that the first mutant strain emerges in the white population (or red population), respectively. 
With high probability we should have $T_0 < \tau_1 < T_1$, which we validated by numerical simulations outlined further on.
As a test of our parameter estimation methodology, we  used  large sets of simulated stochastic genetic trajectories over time durations of 200 days. The simulated stochastic process is based on realistic mutation rates and selective advantages derived from the analysis of the T. Cooper experiments (see \cite{Cooper2001}). For our simulations, the frequency of the event $T_0 < \tau_1 < T_1 <\tau_2$  was of the order of $90\%$ which is a fairly safe accuracy margin in our methodology, as can be seen below.

\section{Estimation of parameters from the first mutant emergence} 
\label{sec:oneevent}
\subsection{Descriptors of the first mutant emergence}
\label{sec5.1}

As outlined in the previous section, for each observed population trajectory, our preliminary estimation algorithm analyzes the observed white cells frequencies $w(t)$ to explicitly compute times $T_0$ and  $T_1$ such that, with very high probability, one has $T_0 <\tau_1 < T_1 < \tau_2$. It is also easy to  determine whether first mutant emergence occurs among white cells or among red cells. Next, we develop more precise estimation algorithm which utilizes observed white cells frequency data $w(t)$ for all
$0 \leq t \leq T_1$ to estimate the three unknowns $\tau_1, \gamma_1, F_{g_1}$ which defines a new genotype $g_1 > 1$.

Denote $\tau_1 $ the random day of  emergence for the first new mutant strain which survives the selection phase on day $\tau_1$ and is, thus, present in the cell population at the beginning of  day $\tau_1 +1$.
Denote  $\gamma_1 >0$ the number of mutants emerging on day $\tau_1+1$. Let $g_1$ be the genotype of these mutants, and denote their unknown growth factor as $F_{g_1}$.

{\bf Notation.}  Define  $(j,w)$-cells and $(j,r)$-cells as white and red cells of genotype $j$, respectively. Denote $w_j(t)$ and $r_j(t)$ the respective frequencies of $(j,w)$-cells and $(j,r)$-cells at the beginning of day $t$. Then, for $0 \leq t \leq \tau_1$ all cells have the same ancestor genotype $1$ so that $w_j(t) = r_j(t) = 0$  for $j \geq 2$ and $w_1(t) = w(t)$, $r_1(t) = r(t)$. 
Recall that our computation of $T_1$, also estimates whether the emerging mutants are white or red cells. These two cases are clearly similar, so without loss of generality we consider that the first emerging mutants are white cells. Analogous formulas can then be immediately derived for the case when emerging mutants are red cells.
To simplify notation we drop the subscript as denote $\tau \equiv \tau_1$ 
and $\gamma\equiv \gamma_1$, since it is clear that we're discussing the first mutation event. 
Let  $\eta$ be the number of white cell mutants present in the terminal population  at the end of day $\tau$ right before the day $\tau$ selection. Then $\eta$  has a Poisson distribution with mean 
$\lambda = N \mu  F_1 w(\tau)$.
%

\subsection{Growth phase}
Since mutants of genotype $g_1$ emerge in the white sub-population 
on day $\tau$ and survive the dilution (selection) step,
at the beginning of day $(1+\tau)$ we have 
\begin{equation} \label{eq:u}
\begin{gathered}
w_1(\tau+1) = w_1(\tau) - {\eta}/N, \\
w_{g_1}(\tau+1) = {\eta}/N, \\
w_j(\tau+1) = 0 \quad \text{for $j \neq 1$ and $j \neq g_1$}, \\
r_1(\tau+1) = r_1(\tau) = 1 - w_1(\tau), \\
r_j(\tau+1) = 0, \quad j = 2,\ldots,g.
\end{gathered}
\end{equation}
For $\tau +1 \leq t < \tau_2$, new mutants may emerge during the day $t$ mutation phase but they necessarily disappear during the day $t$ selection phase because $ t < \tau_2$. Since second mutation occurs on day $t=\tau_2$
and "survives" the dilution step, second mutation event affects population 
only on day $t = \tau_2 + 1$.
Explicit analysis of cell frequencies on day $t = \tau+1$ is presented in \eqref{eq:u}.
Then we consider days $t > \tau+1$ next.
At beginning of day $t$ with  $\tau +1 < t < \tau_2$ the population contains only three cell types,  namely $(1,w)$-cells, $(g_1,w)$-cells, and $(1,r)$-cells, so that
\[
\begin{gathered}
w(t) = w_1(t) + w_{g_1}(t), \\
w_1(t) \geq 0, \; w_{g_1}(t) \geq 0, \; r_1(t) \geq 0, \\
r_j(t) = 0, \quad j=2,\ldots,g,  \\
w_j(t) = 0, \quad \text{for $j \neq 1$ and $j \neq g_1$}, \\
r(t) = r_1(t) = 1 - w(t). \\
\end{gathered}
\]
For $\tau+1 <t < \tau_2$, at the end of the day $t$  growth phase, there are only three non-zero sub-populations with sizes  
\[
\begin{gathered}
\text{number of $(1,w)$-cells} = N  F_1  w_1(t), \\
\text{number of $(g_1,w)$-cells}  = N  F_{g_1}  w_{g_1}(t), \\
\text{number of $(1,r)$-cells } = N  F_1  r_1(t),
\end{gathered}
\]
and all other sub-population sizes are zero.
The total population size at the end of day $t$ growth phase is $N K$, with $K \equiv K(t)$ given by
\begin{equation}
    \label{K}
    K \equiv K(t) = F_1  w_1(t) + F_{g_1}  w_{g_1}(t) + F_1  r_1(t).
\end{equation}

\subsection{Mutations on day $t$  for $\tau +1 \leq t < \tau_2$} 
For days $\tau +1 \leq t < \tau_2$ mutants of new genotypes
can potentially emerge during the mutation step, but these new mutants are necessarily 
eliminated during the day $t$ selection phase, since the second observable 
mutation occurs on day $\tau_2$. However, mutations of genotype 1 into genotype $g_1$
can also occur and might influence the relative composition of 
white and red cells in the population.
Thus, we analyze here the effect of mutations between genotype 1 and genotype $g_1$.

Denote the random number of mutants generated at the end of day $t$ by \\
$\nu_{1,j,w}$ = number of $(1,w)$-cells  mutating  into  $(j,w)$-cells, \\
$\nu_{g_1,j,w}$ = number of $(g_1,w)$-cells  mutating  into  $(j,w)$-cells, \\
$\nu_{1,j,r}$ = number of $(1,r)$-cells  mutating  into  $(j,r)$-cells.

Recall that $\nu_{1,1,w} = \nu_{g_1,g_1,w} = \nu_{1,1,r} = 0$. Moreover, 
$\nu_{1,j,w}$, $\nu_{g_1,j,w}$, and $\nu_{1,j,r}$ are Poisson random variables with 
conditional means and variances as indicated in \eqref{meanmut} so that
\begin{equation}
\begin{aligned}
\bE[\nu_{1,j,w} | \freq(t)] &=Var [\nu_{1,j,w} | \freq(t)] = \mu p_{1 j} N F_1 w_1(t), \\
\bE[\nu_{g_1,j,w} | \freq(t)] &=Var [\nu_{g_1,j,w} | \freq(t)] = \mu p_{g_1 j} N F_{g_1} w_{g_1}(t), \\
\bE[\nu_{1,j,r} | \freq(t)] &=Var [\nu_{1,j,r} | \freq(t)] = \mu p_{1j} N F_1 r_1(t),
\end{aligned}
\end{equation}
where $\freq(t)$ denotes the total histogram of white and red cells.
For $i,j \in  \{1 \ldots g\}$ denote $f_{i,j,w} = \nu_{i,j,w}/N$ and $f_{i,j,r} = \nu_{i,j,r}/N$. 
Since $F_g$ is the largest growth factor, the standard deviations of $f_{i,j,w}$ and $f_{i,j,r}$ verify
\begin{equation}
\label{stdfij}
std(f_{i,j,w}) \leq \sqrt{\mu F_g/N}, \qquad \quad std(f_{i,j,r}) \leq \sqrt{\mu F_g/N}. 
\end{equation}

\subsection{Multinomial Selection on day $t$ for $\tau_1 +1 \leq t < \tau_2$} 
\label{sec5.4}
At the beginning of day $t$ with $\tau +1 \leq t < \tau_2$, there are only three non-zero sub-populations with frequencies
\begin{equation}
\label{freqt}
    freq(t) = [ w_1(t), w_{g_1}(t), r_1(t) ].
\end{equation}
To preserve this property, the selection phase at the end of day $t$ must realize the following event
$\Omega(t)$ = $\{$the random sample of size $N$ selected at the end of  day $t$ contains only three cell types $(1,w)$, $(g_1,w)$, $(1,r)\}$.
Therefore, 
if we denote the population after growth on day $t$ as $POP$, then it is
partitioned into three subpopulations $POP(1,w)$, $POP(g_1,w)$, $POP(1,r)$. 
We denote the population after the mutation step as $MPOP$. 
Denote $Mut(i,j,w)$ the sets of $(i,w)$ cells mutating into $(j,w)$ cells, 
with a similar definition for red mutants $Mut(i,j,r)$. Define 
\[
\begin{gathered}
MPOP(1,w)  = POP(1,w) \cup Mut(g_1,1,w) - \bigcup_{j=1 \ldots g} Mut(1,j,w), \\
MPOP(g_1,w)  = POP(g_1,w) \cup Mut(1,g_1,w) - \bigcup_{j=1 \ldots g} Mut(g_1,j,w), \\
MPOP(1,r)  = POP(1,r) \cup Mut(g_1,1,r) - \bigcup_{j=1 \ldots g} Mut(1,j,r).
\end{gathered}
\]
Define the Restricted Population
$RestPop = MPOP(1,w)  \cup MPOP(g_1,w) \cup MPOP(1,r)$.
For $\tau+1 \leq t < \tau_2$, the event $\Omega(t)$ is necessarily realized, and hence the random sample $\mathcal{S}$ of size $N$ selected at end of day $t$ within $MPOP$ \textit{must be included} in $RestPop$.
Therefore, given  $freq(t)$,  the conditional distribution of genotypes within $\mathcal{S}$ must be a $Multinomial(RestPop,N)$, where one extracts a random sample of size $N$ from the population $RestPop$.
Therefore, changes in $POP$ due to mutations can be quantified by
\begin{equation}
\label{changes}
\begin{gathered}
N H_{1,w} = size(POP(1,w) - MPOP(1,w)), \\
N H_{g_1,w} = size(POP(g_1,w) - MPOP(g_1,w)), \\ 
N H_{1,r} =  size(POP(1,r) - MPOP(1,r)),
\end{gathered}
\end{equation}
which implies
\begin{equation}
\label{Hwr}
N H_{1,w} = \sum\limits_{j=1}^g \nu_{1,j,w} - \nu_{g_1,1,w}, \quad
N H_{g_1,w} = \sum\limits_{j=1}^g \nu_{g_1,j,w} - \nu_{1,g_1,w}, \quad 
N H_{1,r} =   \sum\limits_{j=1}^g   \nu_{1,j,r}.
\end{equation}
The size of $RestPop$ is given by
$size(RestPop) = N (K(t) - H(t))$,  
where
\begin{equation}
    \label{H}
    H=H(t) = H_{1,w} + H_{g_1,w}+ H_{1,r}.
\end{equation}
Due to \eqref{Hwr} and \eqref{stdfij}, the sub-additivity of conditional standard deviation implies   
\begin{equation} \label{stdH}
\begin{aligned}
& std(H_{1,w} | freq(t)) \leq g \sqrt{\mu F_g/N}, \quad 
std(H_{g_1,w} | freq(t)) \leq g \sqrt{\mu F_g/N}, \\
& std(H_{1,r} | freq(t)) \leq g \sqrt{\mu F_g/N}, \quad
std(H | freq(t)) \leq 3 g \sqrt{\mu F_g/N},
\end{aligned}
\end{equation}
since $std(H | freq(t)) = std(H_{1,w} + H_{g_1,w} +H_{1,r})$.
The conditional means are given by
\begin{equation} \label{condmeanHij}
\begin{aligned}
& \bE \big[H_{1,w} | freq(t) \big] = \mu F_1 w_1(t) - \mu p_{g_1,1} F_{g_1} w_{g_1}(t), \\
& \bE \big[H_{g_1,w} | freq(t) \big] = \mu F_{g_1} w_{g_1}(t) -\mu p_{1,g_1} F_1 w_1(t),  \quad
\bE \big[H_{1,r} | freq(t) \big] =  \mu F_1 r_1(t), \\
& \bE \big[H | freq(t) \big] = \mu \big[(1 - p_{1,g_1}) F_1 w_1(t) +(1- p_{g_1,1}) F_{g_1} w_{g_1}(t) 
 + F_1 r_1(t) \big].
\end{aligned}
\end{equation}
The first 3 conditional expectations  are bounded by $\mu F_g$, which in turn implies $\bE \big[H | freq(t) \big] | \leq 3 \mu F_g $. These  bounds are of the order of the mutation rate with typical values 
$\mu \sim 10^{-7}$. Bounds for conditional standard deviations \eqref{stdH} are of the order of $\sqrt{\mu/N} \sim 10^{-6}$ since $N \ge 10^5$.

Right before day $t$ selection, 
frequencies for cell types $(1,w)$, $(g_1,w)$, and $(1,r)$ 
within the sub-population $RestPop$ are given by
\begin{equation}  \label{qwr}
q_{1,w} = \frac{F_1 w_1(t) - H_{1,w}}{K-H}, \quad
q_{g_1,w} = \frac{F_{g_1} w_{g_1}(t)  - H_{g_1,w}}{K-H}, \quad
q_{1,r} = \frac{F_1 r_1(t) - H_{1,r}}{K-H},
\end{equation}
where $K$ is the growth factor for the total population given by \eqref{K} and $H$ is defined in \eqref{H}.
Since with high probability $H/K  = O(10^{-6})$, we introduce the approximation
$$
\frac{1}{(K-H)} \approx 1/K + H/K^2
$$
and, therefore, frequencies in \eqref{qwr} can be approximated as
\begin{equation} \label{Kq.expansion}
\begin{aligned}
K q_{1,w}  & \approx F_1  w_1(t) - H_{1,w} + (H/K)  F_1  w_1(t), \\
K q_{g_1,w} & \approx F_{g_1}  w_{g_1}(t) - H_{g_1,w} + (H/K)  F_{g_1}  w_{g_1}(t),\\
K q_{1,r}  & \approx F_1  r_1(t) - H_{1,r} + (H/K)  F_1  r_1(t).
\end{aligned}
\end{equation}. 

\subsection{Recursive relation to predict $w(t+1)$ at time $t$, for $\tau +1 \leq t < \tau_2$} 
\label{sec5.5}
For $\tau +1 \leq t < \tau_2$ the day $t$ selection must realize the event $\Omega(t)$, which forces the day $t$ multinomial selection to be restricted to the population  $RestPop$. Given the vector $ q = [q_{1,w}, q_{g_1,w}, q_{1,r}]$ of non-zero frequencies in $RestPop$, the multinomial selection in $RestPop$ entails 
$$
\bE[w_1(t+1) | q ] = q_{1,w}, \quad 
\bE[w_{g_1} (t+1) | q ] =q_{g_1,w}, \quad 
\bE[r_1(t+1) | q ] = q_{1,r}.
$$
At time $t$ with $\tau +1 \leq t < \tau_2$, the best predictors $W_1(t+1)$,  $W_{g_1}(t+1)$, $W(t+1)$ of $w_1(t+1)$, $w_{g_1}(t+1)$, $w(t+1)$ are  defined by the conditional expectations
\begin{align*}
& W_1(t+1) =  \bE [w_1(t+1) | freq(t)], \\
& W_{g_1}(t+1) =  \bE [w_{g_1}(t+1) | freq(t)],  \\
& W(t+1) =  \bE [w(t+1) | freq(t)],
\end{align*}
where $freq(t)$ is defined in \eqref{freqt}.
In addition, we have the obvious relation 
$W(t+1) = W_1(t+1) + W_{g_1}(t+1)$.
Since $K\equiv K(t)$ is a deterministic function of $freq(t)$, we then have 
\begin{equation} \label {KWdef}
\begin{aligned}
K(t) W_1(t+1) &= K(t) \bE [w_1(t+1) | freq(t)]  =  \bE [K(t)  q_{1,w} | freq(t)], \\
K(t) W_{g_1} (t+1) &= K(t) \bE [w_{g_1} (t+1) | freq(t)] =\bE [K(t) q_{g_1,w} | freq(t)] .
\end{aligned}
\end{equation}
From \eqref{KWdef} and \eqref{Kq.expansion} we then obtain
\begin{equation}\label{lastKW}
\begin{aligned}
K(t) W_1(t+1) & \approx 
F_1 w_1(t) - \bE \big[H_{1,w} | freq(t) \big] + (F_1/K) w_1(t) \bE \big[H | freq(t) \big],  \\
K(t) W_{g_1}(t+1) & \approx 
F_{g_1}  w_{g_1} (t) - \bE \big[H_{g_1,w} |  freq(t) \big] + (F_{g_1} /K)  w_{g_1} (t) \bE \big[H | freq(t) \big],
\end{aligned}
\end{equation}
and due to \eqref{condmeanHij}, \eqref{lastKW}
\begin{equation}\label{kw1a}
\begin{aligned}
K(t) W_1(t+1) & \approx F_1 w_1(t)  (1- \mu + \mu Z /K ) + \mu p_{g_1,1} F_{g_1} w_{g_1}(t), \\
K(t) W_{g_1}(t+1) & \approx F_{g_1} w_{g_1}(t)  (1-\mu + \mu Z /K ) + \mu p_{1,g_1} F_1 w_1(t) ,
\end{aligned}
\end{equation}
where
\[
\begin{aligned}
Z\equiv Z(t) & =  (1- p_{1,g_1}) F_1 w_1(t) +(1- p_{g_1,1}) F_{g_1} w_{g_1}(t) + F_1 r_1(t) ,\\
K\equiv K(t) & = F_1  w_1(t) + F_{g_1}  w_{g_1}(t) + F_1  r_1(t).
\end{aligned}
\]
In equation \eqref{kw1a} the only two unobserved freqencies are $w_1(t)$ and $w_{g_1}(t) = w(t) - w_1(t)$, since $r_1(t) =r(t)$ is actually observed. The unobserved $w_1(t)$ can be approximated by its one-step conditional mean $W_1(t) = \bE[w_1(t) | freq(t-1)]$,  and  $w_{g_1}(t)$  is then approximated by $w(t) - W_1(t)$. We thus obtain the following recurrence relation between  $W_1(t)$ and $W_1(t+1)$, as well as an evaluation of $W_{g_1}(t+1)$
\begin{equation}\label{kw1b}
\begin{aligned}
K(t) W_1(t+1) & \approx F_1 W_1(t) (1 - \mu + \mu z(t)/k(t) )+  \mu p_{g_1,1}  F_{g_1} (w(t) -W_1(t)), \\
K(t) W_{g_1} (t+1) & \approx F_{g_1} (w(t) - W_1(t)) (1- \mu + \mu z(t)/k(t)  ) + \mu p_{1,g_1} F_1 W_1(t),
\end{aligned}
\end{equation}
where $Z(t) \approx z(t)$ and $K(t) \approx k(t)$ with
\[
\begin{aligned}
z(t) & = (1 - p_{1,g_1}) F_1 W_1(t)  + (1 - p_{g_1,1}) F_{g_1} (w(t) -W_1(t))+ F_1 (1-w(t)), \\
k(t) & = F_1 + (F_{g_1} - F_1) (w(t) - W_1(t)).
\end{aligned}
\]
Therefore, 
The best predictor $W(t+1)$ of $w(t+1)$ verifies 
$W(t+1) = W_1(t+1) + W_{g_1}(t+1)$ and is hence given by
\begin{equation}\label{predictW}
\begin{aligned}
K(t) W(t+1) & \approx (1 - \mu + \mu z(t)/k(t) ) \times \big[ (F_1 - F_{g_1}) W_1(t)
 F_{g_1} w(t) \big] \\
 & +  \mu \big[ p_{g_1,1} F_{g_1} w(t) + (p_{1,g_1} F_1 - p_{g_1,1}  F_{g_1} )W_1(t) \big].
\end{aligned}
\end{equation}

\subsection{Non-linear least squares estimates of $\tau_1, \gamma_1, s_{g_1}$}
\label{sec5.6}

As discussed in section \ref{sec5.1}, we first analyze each trajectory which contains
daily frequencies of the white sub-population
$\{ w(t), \, t=1,2, ....,\}$ and estimate time-interval
$RTIME = [T_0, T_1]$ containing with very high probability 
the unknown emergence time $\tau_1$ of the first mutation.

In standard laboratory experiments for Escherichia Coli genetic evolution (see \cite{Cooper2001}), the growth factor $F_1$ of the ancestor cells is well known, and the mutation rate $\mu$ can often be considered as already known from long term past experiments. One then  seeks to estimate the selective advantage $ s_{g_1}$ of the first mutation event with unknown genotype $g_1$. This selective advantage defines the growth factor $F_{g_1}=(F_1)^{1+s_{g_1}}$. To develop an optimization procedure,
one first needs to have an initial estimate for a plausible range
for the unknown selective advantage. For instance, we take this range to be
$RSEL = [1\% , 20\%]$. Moreover, we also need to have an estimate for a  plausible 
range $RMUT$ for the initial (unknown) number of new emerging mutants present at 
the beginning of  day $\tau_1 +1$. For Escherichia Coli experiments with $N= 10^5$,  
one can for instance take $ RMUT = [1, 500]$.

We define an optimization procedure and perform a grid search over all possible 
combinations of
$(\tau, s, \gamma) \in RTIME \times RSEL \times RMUT$. 
We use the triplet $(\tau, s, \gamma)$ as a candidate for optimal values
$(\tau_1, s_{g_1}, \gamma_1)$ in formula \eqref{kw1b}
in the previous section.
Given triplet $(\tau, s, \gamma)$ and white population $w(t)$ at time $t$, 
the best predictor of the white population on the next day, $w(t+1)$, is 
$W(t+1) = \bE[w(t+1) | freq(t)]$ and the associated Squared Error is
$SE= (w(t+1) - W(t+1))^2$.

Recall, that $\tau$ is the estimator for the time of the first mutation event.
Therefore, all cells have the same (ancestor) genotype until mutation and, thus, for $t < \tau$
\[
W(t+1) = w(t) \qquad \text{and} \quad 
SE(t) = (w(t+1) - w(t))^2 \qquad \text{for all} \quad T_0 \le t < \tau.
\]
For $t=\tau$, since $\gamma$  mutants emerge on day $\tau+1$, we can approximate $W(\tau + 1)$ by 
$$
W(\tau+1) = w(\tau) + \gamma/N \qquad \text{and} \qquad  SE(\tau) = (w(\tau+1) - w(\tau) -  \gamma / N)^2.
$$
For $\tau+1 \leq t  \leq T_1$, we can recursively compute $W(t+1) $ from the first formula  in \eqref{kw1b}, and this directly  gives us the  associated squared errors of prediction
\[
SE(t) = (w(t) - W(t))^2 \quad \text{for} \quad  \tau+2 \leq t \leq T_1.
\]
The Mean Squared Error is then computed as
$$
MSE(\tau, s, \gamma) = \frac{1}{T_1-T_0} \sum_{t= T_0}^{T_1} SE(t)
$$
and the best estimate for the pair $(\tau_1, s_{g_1})$ is computed 
by minimizing the $MSE$
\[
(\hat{\tau}_1, \hat{s}_{g_1}, \hat{\gamma}_1) = \argmin_{\tau, s, \gamma} \frac{1}{T_1-T_0} \sum_{t= T_0}^{T_1} SE(t).
\]
We solve the optimization procedure above by a straightforward multi-scale grid search over the set $RTIME \times  RSEL \times RMUT$.

Theefore, 
\textit{each observed trajectory} provides three estimates 
$(\hat{\tau}_1, \hat{s}_{g_1}, \hat{\gamma}_1)$
for the first mutants emergence time,  for the selective advantage of these first mutants, and for the number of new mutants born on day $\tau_1$ and present at the beginning of day $\tau_1 + 1$, respectively.
This optimization procedure also provides an estimate $\hat{F}_{g_1}$ for the growth factor $F_{g_1}$. 
In Section {sec:pool} we explain how we pool estimates for each individual trajectory 
into a global estimate over the whole ensemble.


%

\section{Estimation of $T_2$ such that $ \hat{\tau}_1 < \tau_2 \leq T_2$}
\label{sec:T2}

In previous sections we discussed how to use 
a single observed trajectory of white cell frequencies $w(t)$ to estimate 
parameters $\hat{\tau}_1$, $\hat{s}_1$, $\hat{\gamma}_1$, $\hat{F}_{g_1}$ 
describing the first mutation event. 
After computing these estimates,  we can use equations \eqref{kw1b}
iteratively to compute estimates for the evolution of the white sub-population until the second mutation event, i.e. we can use \eqref{kw1b}  to compute $W_1(t)$ and $W_{g_1} (t)$ of $w_1(t)$  and $w_{g_1}(t)$ for $\hat{\tau}_1 \leq t \le \tau_2$. 

Triple $(\hat{\tau}_1, \hat{s}_1, \hat{\gamma}_1)$ can be used to compute the 
behavior of mutants and estimate the frequency of the white sub-population until the second mutation. Estimates for the white sub-population are given by
$W(t) = W_1(t) + W_{g_1} (t)$ for $\hat{\tau}_1 < t < \tau_2$ (here $\tau_2$ is unknown). 
On the other hand, 
observational data for the frequency of the white sub-population is
also available, i.e. we observe $w(t)$ for $\hat{\tau}_1 < t < \tau_2$.
Therefore, we can compute the accuracy of prediction for $\tau_1 < t < \tau_2$ as $SE(t)= (w(t+1)-W(t+1))^2$ and define the 
moving mean squared error of prediction 
$$
MMSE(t) = \frac{1}{t-\hat{\tau}_1}\sum_{s= \hat{\tau}_1}^{t} SE(s).
$$ 
It is reasonable to assume that $ MMSE(t)$ remains reasonably small
until the second mutation occurs, i.e. for $t < \tau_2$, 
but increases rapidly for $t > \tau_2$ (after the second mutation).
This allows us to find an interval for the second mutation. In particular, we
define $T_2$ to be the first day $t> \hat{\tau}_1 $ such that for $t \leq u \leq t+2$ one has  $SE(u) > 2.5 MMSE(t)$ and $(w(u) - W(u))$ has the same sign for $u=t$, $t+1$, and $t+2$. 
If the sign of $w(t) - W(t)$ is positive (negative), then the second strain of new mutants  emerges in the white (red) sub-population, respectively.
We can also conclude that with high probability the second mutation occurs 
in the time interval $[\hat{\tau}_1, T_2]$.

\section{Non-Linear least squares applied to the second mutant emergence}
\label{sec:twoevent}

In section \ref{sec:oneevent} we discussed to how to estimate 
parameters $\tau_1$, $\gamma_1$, $s_{g_1}$ from the first mutation event.
This also allows to compute the corresponding growth factor,
$F_{g_1}$, of the new strain.
Without loss of generality we can assume that the first mutation event 
occurs in the white sub-population. However, the second mutation event might
occur in the sub-population of the same (white) or opposite (red) color
and we have to analyze both cases.
Denote $\tau_2$ the time of the second mutation event, $\gamma_2$ the number of new mutants present at beginning of day $\tau_2 + 1$, and $g_2$  the genotype of new mutants. 
The probabilistic analysis of the second mutant emergence is quite similar to the analysis outlined in section \ref{sec:oneevent}.

\subsection{Case 1: the first and second mutation events occur in sub-population of the same color}
\label{twosame}
Without loss of generality we assume that both the first and second mutant strains emerge among white cells. Thus (if $\tau_3$ is the time of the third mutation or the end of experiment), 
for $\tau_2 < t <\tau_3$, at beginning of day $t$, 
the only non-zero genotype frequencies are 
$\{w_1(t)$, $w_{g_1}(t)$, $w_{g_2}(t)$, $r_1(t)\}$. 
Hence we have
\begin{equation}
\label{linear}
r(t) = r_1(t) = 1 - w(t) , \quad 
w_{g_2}(t) = w(t) - w_1(t) - w_{g_1}(t).
\end{equation}
Therefore we define vector
\begin{equation}
    \label{freq2}
    freq(t) = \left[ w(t), w_1(t), w_{g_1}(t), w_{g_2}(t) \right],
\end{equation}
which
contains all the information about frequencies of 
four sub-populations on day $t$.

Given $freq(t)$, 
appropriate predictors for frequencies of white sub-populations
$w_1(t+1)$, $w_{g_1}(t+1)$, $w_{g_2}(t+1)$, are given by
\begin{align*}
W_1(t+1) & = E[ w_1(t+1) | freq(t) ], \\
W_{g_1}(t+1) & = E[ w_{g_1} (t+1) | freq(t) ], \\
W_{g_2}(t+1) & = E[ w_{g_2} (t+1) | freq(t) ].
\end{align*}

Similar to section \ref{sec5.5}, our goal is to obtain explicit formulas for 
conditional predictors for white frequency in the population.
Thus, we can apply the same 
probabilistic arguments as in section \ref{sec5.5} to derive the following
formulas for frequencies of white sub-populations for all three genotypes.
In particular, we obtain
\begin{equation}
\label{same2}
\begin{aligned}
K(t) W_1(t+1) & \approx F_1 w_1(t) - \mu x(t) + (\mu/K(t)) \, F_1 w_1(t)  \, X(t), \\
K(t) W_{g_1}(t+1) &\approx 
 F_{g_1} w_{g_1}(t) - \mu y(t)  + (\mu/K(t)) \, F_{g_1} w_{g_1}(t) \, Y(t), \\
K(t) W_{g_2}(t+1) &\approx  F_{g_2} w_{g_2}(t) - \mu z(t) + (\mu/K(t)) \, F_{g_2} w_{g_2}(t) \, Z(t),
\end{aligned}
\end{equation}
where
\begin{equation*}
\begin{aligned}
x(t) &= F_1 w_1(t) - p_{g_1, 1}F_{g_1} w_{g_1}(t) - p_{g_2, 1}F_{g_2} w_{g_2}(t), \\
X(t) &= (1 - p_{1,g_1} - p_{1, g_2}) F_1 w_1(t)  + (1 - p_{g_1, 1}) F_{g_1} w_{g_1}(t)
+ (1 - p_{g_2, 1}) F_{g_2} w_{g_2}(t)  +  F_1 r_1(t) ,
\end{aligned}
\end{equation*}
\begin{equation*}
\begin{aligned}
y(t) &= F_{g_1} w_{g_1}(t) - p_{1,g_1}F_1 w_1(t) - p_{g_2,g_1}F_{g_2} w_{g_2}(t), \\
Y(t) &= (1 - p_{1,g_1} - p_{1,g_2}) F_1 w_1(t)  + (1 - p_{g_1,1}) F_{g_1} w_{g_1}(t) 
+ (1 - p_{g_2,1}) F_{g_2} w_{g_2}(t)  +  F_1 r_1(t) ,
\end{aligned}
\end{equation*}
\begin{equation*}
\begin{aligned}
z(t) &= F_{g_2} w_{g_2}(t) - p_{1,g_2}F_1 w_1(t)- p_{g_1,g_2}F_{g_1} w_{g_1}(t), \\
Z(t) &= (1 - p_{1,g_1} - p_{1,g_2}) F_1 w_1(t)  + (1 - p_{g_1,1}) F_{g_1} w_{g_1}(t) 
+ (1 - p_{g_2,1}) F_{g_2} w_{g_2}(t)  +  F_1 r_1(t) .
\end{aligned}
\end{equation*}
The term $K(t)$ in equations \eqref{same2} is  given by 
$$
K(t) = F_1 w_1(t) + F_{g_1} w_{g_1}(t) + F_{g_2} w_{g_2}(t) + F_1 r_1(t)
$$
and due to the linear relations \eqref{linear}, $K(t)$ can be expressed as
\begin{equation}
\label{same4}
K(t) = F_1 + (F_{g_2} - F_1) (w(t) - w_1(t)) - (F_{g_2} - F_{g_1}) w_{g_1}(t).
\end{equation}

Formulas  \eqref{same2} and \eqref{same4} 
define an explicit vector valued function $\Phi : R^3 \to R^3$ such that
\begin{equation}
\label{iter}
[ W_1(t+1), W_{g_1}(t+1), W_{g_2}(t+1) ] =  \Phi (w_1(t), w_{g_1}(t), w_{g_2}(t) ),
\end{equation}
providing optimal predictions for frequencies of
white sub-populations on day $t+1$ given those frequencies on day $t$.

Since the second mutation occurs on day $t=\tau_2$ and becomes "visible" on day 
$t=\tau_2+1$, for $\hat{\tau}_1 \leq t \leq \tau_2$, we have already computed (see section \ref{sec5.5}) estimates $W_{1}(t)$ and $W_{g_1}(t)$ of frequencies of white sub-populations $(1,w)$ and $(g_1,w)$. 
Combining two sub-populations 
provides an estimate $W(t) = W_{1}(t) + W_{g_1}(t)$ for the frequency of white cells for $\hat{\tau}_1 \leq t \leq \tau_2$.
Since second mutation occurs at $t=\tau_2$, we assume that $w_{g_2}(\tau_2)= 0$, and
substituting $W_{1}(\tau_2)$ and $W_{g_1}(\tau_2)$ into \eqref{iter}
to obtain estimates for $W_1(\tau_2+1), W_{g_1}(\tau_2+1)$. 
Next, we assume that $\gamma_2$ mutants emerge during the mutation and, thus, the optimal estimate for the frequency of $(g_2, w)$ cells on day $\tau_2+1$ is
$W_{g_2}(\tau_2+1) = \gamma_2/N$.
After that, for $ 2+\tau_2  \leq t \leq T_2$ we expect our estimators $W_1(t)$, $W_{g_1}(t)$, $W_{g_2}(t)$ to be quite close to frequencies 
of $(1,w)$, $(g_1,w)$, and $(g_2,w)$ sub-populations since no mutations occur during that time-interval.
Therefore, we can compute optimal estimates for frequencies of white sub-populations
for the three genotypes
iteratively by the recurrence formula
$$
[W_1(t+1), W_{g_1}(t+1), W_{g_2}(t+1) ] =  \Phi \left(W_1(t), W_{g_1}(t), W_{g_2}(t) \right).
$$

Combining estimates above, we obtain estimates for the frequency of white cells in the population
$W(t) = W_1(t) + W_{g_1}(t) + W_{g_2}(t)$ for $\tau_2 + 1  \leq t \leq T_2$. 
Combing the above expression with predictor 
$W(t)$ for $\hat{\tau}_1 + 1  \leq t \leq \tau_2$ derived in section 
\ref{sec5.5}, we obtain explicit formulas for computing the optimal predictor of white-cell frequency
$W(t)$ for $\hat{\tau}_1 +1 \leq t \leq T_2$.
Therefore, the associated mean squared error of prediction can be computed as
$$
MSE = \frac{1}{T_2 - \hat{\tau}_1 } \sum_{t = \hat{\tau}_1 + 1}^{T_2}  (W(t) - w(t))^2 .
$$

Of course the $MSE$ above depends on the three unknown values $\tau_2, s_{g_2}, \gamma_2$, which we want to  estimate. Similar to discussion in section \ref{sec5.6}, we can define 
a non-linear least squares approach which consists in minimizing $MSE(\tau_2, s_{g_2}, \gamma_2)$ over all triplets $(\tau_2, s_{g_2}, \gamma_2)$ where   $s_{g_2} \in RSEL, \gamma_2 \in RMUT$ and 
$\tau_2 \in \left[\hat{\tau}_1+1 ,  T_2 \right]$. Similar to section \ref{sec5.6} 
this minimization is done by multigrid search and is reasonably fast 
for the case of four genotypes. The minimizing triplet provides the least squares estimates 
$(\hat{\tau}_2, \hat{s}_{g_2}, \hat{\gamma}_2)$.

\subsection{Case 2: the first and second mutation events occur in sub-populations of distinct colors}
\label{twodiff}

In this section we derive formulas for optimal prediction of white- and red-cell frequencies when
the first and second mutations occur in sub-populations of different colors. 
Since the approach used here is analogous to the discussion in sections \ref{sec5.5} and 
\ref{twosame}, we only present the final result.

Denote conditional expectations of $w_1(t+1)$, $w_{g_1}(t+1)$, $r_1(t+1)$, $r_{g_2}(t+1)$ 
given frequencies $[w_1(t), w_{g_1}(t), r_1(t), r_{g_2}(t)]$ as
$W_1(t+1)$, $W_{g_1}(t+1)$, $R_1(t+1)$, $R_{g_2}(t+1)$.
Thus, we obtain the following four formulas valid for $\tau_1+1 \leq t \leq T_2$
\begin{equation}
\label{diff2}
\begin{aligned}
   K(t) W_1(t+1) &\approx  F_1 w_1(t) - \mu a(t)+ (\mu/K(t)) \, F_1 w_1(t) \, A(t) , \\
   K(t) W_{g_1}(t+1) &\approx  F_{g_1} w_{g_1}(t) + \mu b(t) + (\mu/K(t)) \, F_{g_1} w_{g_1}(t) \, B(t) , \\
   K(t) R_1(t+1) &\approx  F_1 r_1(t) - \mu c(t) + (\mu/K(t)) \, F_1 r_1(t) \, C(t), \\
   K(t) R_{g_2} (t+1) &\approx  F_{g_2} r_{g_2} (t) + \mu d(t) + (\mu/K(t)) \, F_{g_2} r_{g_2}(t) \, D(t),
\end{aligned}
\end{equation}
where
\begin{equation*}
\begin{aligned}
a(t)  &= F_1 w_1(t) - p_{g_1,1} F_{g_1} w_{g_1}(t), \\
A(t) &=  (1-p_{1,g_1})F_1 w_1(t) + (1-p_{g_1,1})F_{g_1} w_{g_1}(t) +(1-p_{1,g_2})F_1 r_1(t) + (1-p_{g_2,1})F_{g_2} r_{g_2}(t),
\end{aligned}
\end{equation*}
\begin{equation*}
\begin{aligned}
b(t) &= p_{1,g_1}F_1 w_1(t)-F_{g_1} w_{g_1}(t), \\
B(t) &= (1-p_{1,g_1})F_1 w_1(t) + (1-p_{g_1,1})F_{g_1} w_{g_1}(t) +(1-p_{1,g_2})F_1 r_1(t) + (1-p_{g_2,1})R_{g_2} r_{g_2}(t),
\end{aligned}
\end{equation*}
\begin{equation*}
\begin{aligned}
c(t) &= F_1 r_1(t) - p_{g_2,1}F_{g_2} r_{g_2}(t),  \\
C(t) &= (1-p_{1,g_1})F_1 w_1(t) + (1-p_{g_1,1})F_{g_1} w_{g_1}(t) +(1-p_{1,g_2})F_1 r_1(t) 
+ (1-p_{g_2,1})F_{g_2} r_{g_2}(t),
\end{aligned}
\end{equation*}
\begin{equation*}
\begin{aligned}
d(t) &= p_{1,g_2}F_1 r_1(t)-F_{g_2} r_{g_2}(t), \\
D(t) &= (1-p_{1,g_1})F_1 w_1(t) + (1-p_{g_1,1})F_{g_1} w_{g_1}(t) +(1-p_{1,g_2})F_{g_2} r_{g_2}(t) + (1-p_{g_2,1})F_{g_2} r_{g_2}(t).
\end{aligned}
\end{equation*}
The term $K(t)$
in equations \eqref{diff2} is given by
$K(t) = F_1  w_1(t) + F_{g_1}  w_{g_1}(t) + F_1  r_1(t) + F_{g_2} r_{g_2}(t)$
and since we have
$w_{g_1}(t) = w(t) - w_1(t)$ and $r_{g_2}(t) = r(t) - r_1(t) = (1 - w(t)) - r_1(t)$,
$K(t)$ becomes
\begin{equation}
\label{diff5}
K(t)  = F_{g_2} - (F_{g_1}-F_1) w_1(t) - (F_{g_2}-F_1) r_1(t) - (F_{g_2}-F_{g_1}) w(t).
\end{equation}

Similar to section \ref{twosame} we can apply equations \eqref{diff2} 
to compute iteratively values of $W_1(t)$, $W_{g_1}(t)$, $R_1(t)$, $R_{g_2}(t)$, 
for $\tau_1 \leq t \leq T_2$. 
Since these terms are optimal predictors of white- and red-cells frequencies for different genotypes, the optimal predictor for the total frequency of the white-cell sub-population can be computed as
$W(t) = W_1(t) + W_{g_1}(t)$. Note, that the frequency of the white cells depends on the 
frequency of $(g_2,r)$ cells, $R_{g_2}$, through $A(t)$ and $B(t)$ defined above.
Therefore, similar to section \ref{twosame}, we, can define the mean squared error and the corresponding
optimization procedure to compute the estimates $(\hat{\tau}_2, \hat{s}_{g_2}, \hat{\gamma}_2)$.

\section{Pooling estimates for the selective advantages of mutants}
\label{sec:pool}

\subsection{Multiple observed trajectories}
In a typical evolutionary experiment, many colonies of the same bacteria
(starting with the same ancestor genotype) evolve in parallel in a "rack" of distinct wells. All colonies undergo similar daily cycles of growth-mutations-selection. However, mutations do not occur in these
colonies simultaneously. Therefore, such a multi-well long term experiment generates a finite number of different observed trajectories 
for white cell frequencies $w(t)$. While all colonies start with the same 
ancestor genotype, white-cell frequencies for later times might differ
for different colonies depending on a particular mutation which occurred in each colony.
Observed long trajectories which exhibit zero emergence of new mutants strain generally 
constitute a  small percentage of trajectories. 
As outlined above, we can automatically detect trajectories $\omega_1, \ldots, \omega_n$ 
having at least one emergence of a new mutant strain carrying a beneficial mutation. 
Next, for each observed trajectory $\omega_j$, we can apply our non-linear least squares estimation algorithm to generate an estimate $\hat{s}(j)$ for the selective advantage of the first mutation event. 
The next step is to 
\emph{combine these $n$ estimates}. 
To this end, we treat individual estimates $\hat{s}(1)$, $\hat{s}(2)$,$\ldots$, $\hat{s}(n)$ as independent observations and
combine them into a histogram $H_1$. Since in our model
genotypes with higher selective advantage constitute a finite list of size $g-1$, 
we expect histogram $H_1$ to exhibit at most $g-1$ local peaks
centered around (unknown) $s_2, \ldots , s_g$.

Of course we also apply our non-linear least squares analysis to the second emergence of a new mutant strain. This is done for $m \leq n$ trajectories exhibiting at least two new mutant strains, and provides another histogram $H_2$ of $m$ estimates of selective advantages. Similar to the histogram $H_1$ discussed above, histogram $H_2$ should exhibit  at most $g-1$ local peaks centered around $s_2, \ldots, s_g$.

We have avoided mixing blindly the two histograms $H_1$ and $H_2$ because the frequency peaks in $H_2$ are generally wider and hence less precisely centered than the peaks in $H_1$. Thus, it is more efficient to first compute the center and dispersion of each peak in $H_1$ to generate first estimators of $s_2, s_3, \ldots, s_g$. These estimators and their accuracy can then be boosted by a similar detection of new frequency peaks present in $H_2$. This process can be extended to the analysis of the third emergence of mutant strains, to generate a histogram $H_3$ of selective advantages estimates, and so on. 
However, the accuracy of frequency peaks in $H_3$ is typically weaker than in $H_2$. One of the reasons for this is that the number of trajectories exhibiting three successive mutation events is typically much smaller than $m$. Therefore, when we tested this approach on  simulated trajectories, we restricted the histogram analysis to the first two mutations and histograms $H_1$ and $H_2$.
%

\subsection{Gaussian Mixture Fitting to detect  Histogram Peaks }
\label{peaks}

The histogram $H_1$ of $n$ selective advantages described in the previous section 
typically has $g-1$ peaks of frequencies. To detect them we fit a mixture  of $g-1$ 
Gaussians to this histogram. In particular, we consider a density function of the form
\begin{equation}
\label{multiG}
G(x) = \sum\limits_{i=2}^{g} a_i e^{-\frac{(x - b_i)^2}{c_i^2}}
\end{equation}
where $a_i, b_i, c_i$ are unknown positive parameters with the restriction 
$ 0 < b_2 < b_3 < \ldots < b_{g}$ since $b_i$ are the estimates for the selective advantages.
Thus, there are $3(g-1)$ parameters to optimize. 

We fit $G(x)$ to the histogram $H_1$ by a standard non-linear least squares algorithm. This technique  provides an acceptable fit when the number of parameters verifies $3(g-1) \ll n$.
Least squares fitting provides estimates $\hat{a}_i$, $\hat{b}_i$, $\hat{c}_i$  with theoretical accuracy of the order of $1/\sqrt{n- 3(g-1)}$. 
The $g-1$ estimated peak centers $\hat{b}_2, \ldots, \hat{b}_g$ are then our estimates $\hat{s}_2, \ldots, \hat{s}_g$ of the unknown selective advantages. Each estimated standard deviation $\hat{c}_i$ provides an approximate measure of the error for the corresponding estimate $\hat{s}_i$.

In practice, the number of genotypes $g$ is not really known a priori, 
and the analysis of histogram $H_1$ enables a first estimation of $g$ as the number of 
significant estimated peaks in $H_1$. Moreover, analysis of the histogram $H_1$ provides 
a set of estimates for the selective advantages of these genotypes. 
Subsequent analysis of the second histograms $H_2$ provides a second set of estimates
for the number of genotypes and their selective advantages.
A statistical comparison between the first (based on $H_1$) the second (based on $H_2$) sets of estimates can then determine whether $H_2$ has peaks significantly different from the $H_1$ peaks, and improve (or validate) the estimation of $g$. After this comparison, the pairs of peaks in $H_1$ and $H_2$ which have significantly matching centers can then be combined to provide a more accurate estimation of selective advantages.

\section{Results}
\label{sec:sims}

In this paper we focus on validating the applicability and accuracy of the
estimation approach using large synthetic datasets of simulated trajectories.
In particular, we consider a model where every mutation occurrence in the population is 
equally likely to generate any one of the stronger mutants. 
In this case, the transition matrix $\mathcal{P}$ becomes
\begin{equation}
\label{P}
\mathcal{P} = \left( \begin{array}{ccccc}
0 & \frac{1}{g-1} &\frac{1}{g-1} & \cdots & \frac{1}{g-1} \\
0 & 0 & \frac{1}{g-2} & \cdots &\frac{1}{g-2}\\
\vdots & & \ddots & & \vdots\\
0 & 0 & 0 & \cdots & 1\\
0 & 0 & 0 & \cdots & 0
\end{array} \right) .
\end{equation}
Therefore, entries of the matrix $\mathcal{P}$ in our model are given by 
\begin{equation} 
\label{pij}
p_{ij} =
\begin{cases} 
\frac{1}{g-i} & i < j  \\
0 & \text{otherwise}.
\end{cases}
\end{equation}

To validate our approach we generate a fixed number of trajectories using the above model
with particular value of the mutation rate, $\mu$, and selective advantages $s_2$, $s_3$, $\ldots$, $s_g$. We then apply our estimation algorithm to these simulated trajectories to recover estimates 
for the selected advantages $\hat{s}_2$, $\hat{s}_3$, $\ldots$, $\hat{s}_g$.

\subsection{Parameters of the simulated process}
\label{sec:params}
Parameters of the stochastic model for population growth are
(i) the number of  genotypes is $g= 4$;
(ii) the number of simulated trajectories is  $n= 1000$; 
(iii)  $N = 50000$ is the daily initial population size;
(iv)  on the first day all cells have the same ancestor genotype with growth factor $F_1 = 200$; 
(v)  on the first day $N/2$ cells are white cells and $N/2$ cells are red cells;
(vi) cell colors are preserved during growth, mutation, and selection;
(vii) mutant genotypes have selective advantages $s_2=0.04$, $s_3=0.07$, $s_4=0.13$;
(viii) the mutation rate is $\mu = 10^{-6}$.

Selective advantages above correspond to growth factors
$(F_2, F_3, F_4)=(247, 290, 398)$.
These parameter values are consistent with numerical values reported in past and ongoing experiments for Escherichia Coli (see e.g. \cite{Gordo2012, Zhang2012}.
We also fix  the  mutation transition matrix, as indicated in \eqref{pij} which stipulates that only beneficial mutations can occur. 

We simulated $n = 1000$ random trajectories with uniform length $T=200$ days. 
At the beginning of each day $t \in [1, 200]$, we record the total white cells frequency $w(t)$ 
as well as the eight frequencies  $w_j(t)$, $r_j(t)$, $j=1,\ldots,4$
corresponding to sub-populations $j$-white or $j$-red cells.
Of course, these detailed eight frequencies are not observable in actual experimental data, where only total frequencies of the white and red sub-populations are observed. Therefore, the eight unobservable frequencies are not used in computations of estimates for selective advantages.
Nevertheless, frequencies of white and red sub-populations for each genotype 
provide additional data for validating our approach and, in particular, 
for validating estimation of intervals for the 
first and second mutations discussed in sections \ref{sec:T1} and \ref{sec:T2}, respectively.

\begin{figure}[h]
	\centerline{\includegraphics[width=3 in, height = 1.8 in]{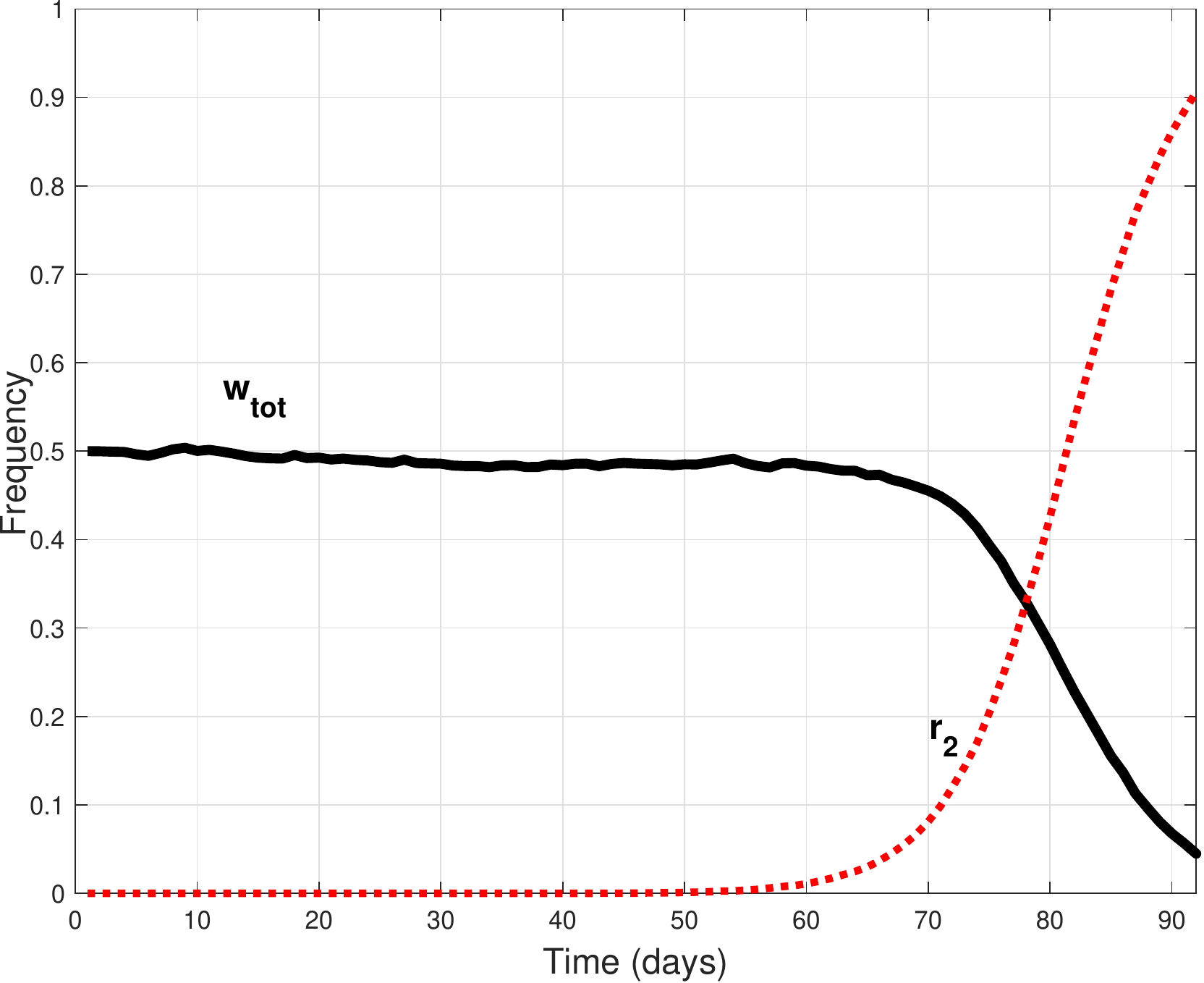}
	\includegraphics[width=3 in, height = 1.8 in]{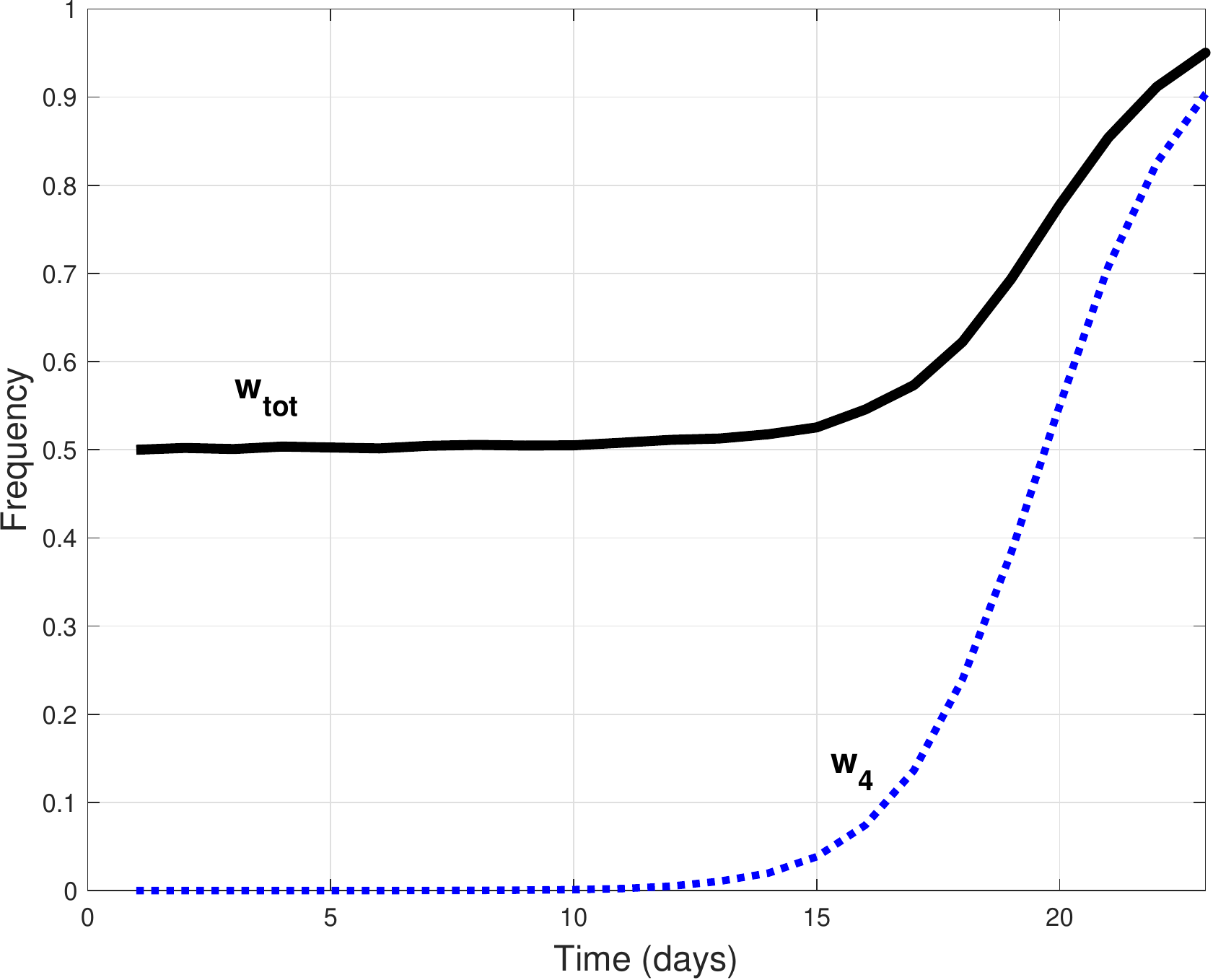}}
	\caption{Two examples of trajectories with only one mutation event. 
	We	display the frequencies of white cells and of the sub-population of mutants.
	Black line (labeled $w_{tot}$) = total white cells frequency $w(t)$, 
	Left part - emergence of genotype-$2$ mutants among red cells (red line $r_2$), 
	Right part - emergence of  genotype-$4$ mutants among white cells (blue line $w_4$).}
	\label{fig3}
\end{figure}

\begin{figure}[h]
	\centerline{\includegraphics[width=3 in, height = 1.8 in]{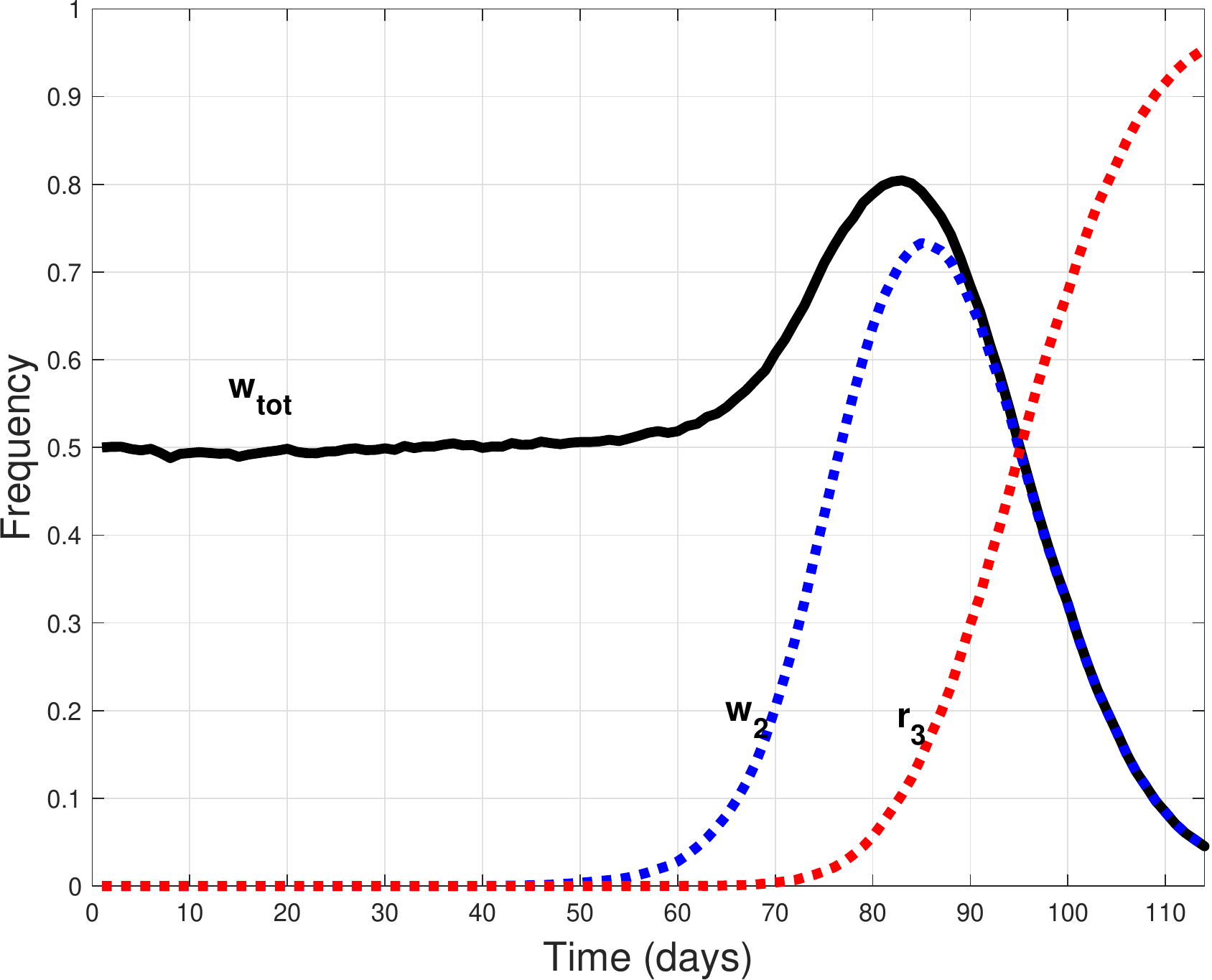}
	\includegraphics[width=3 in, height = 1.8 in]{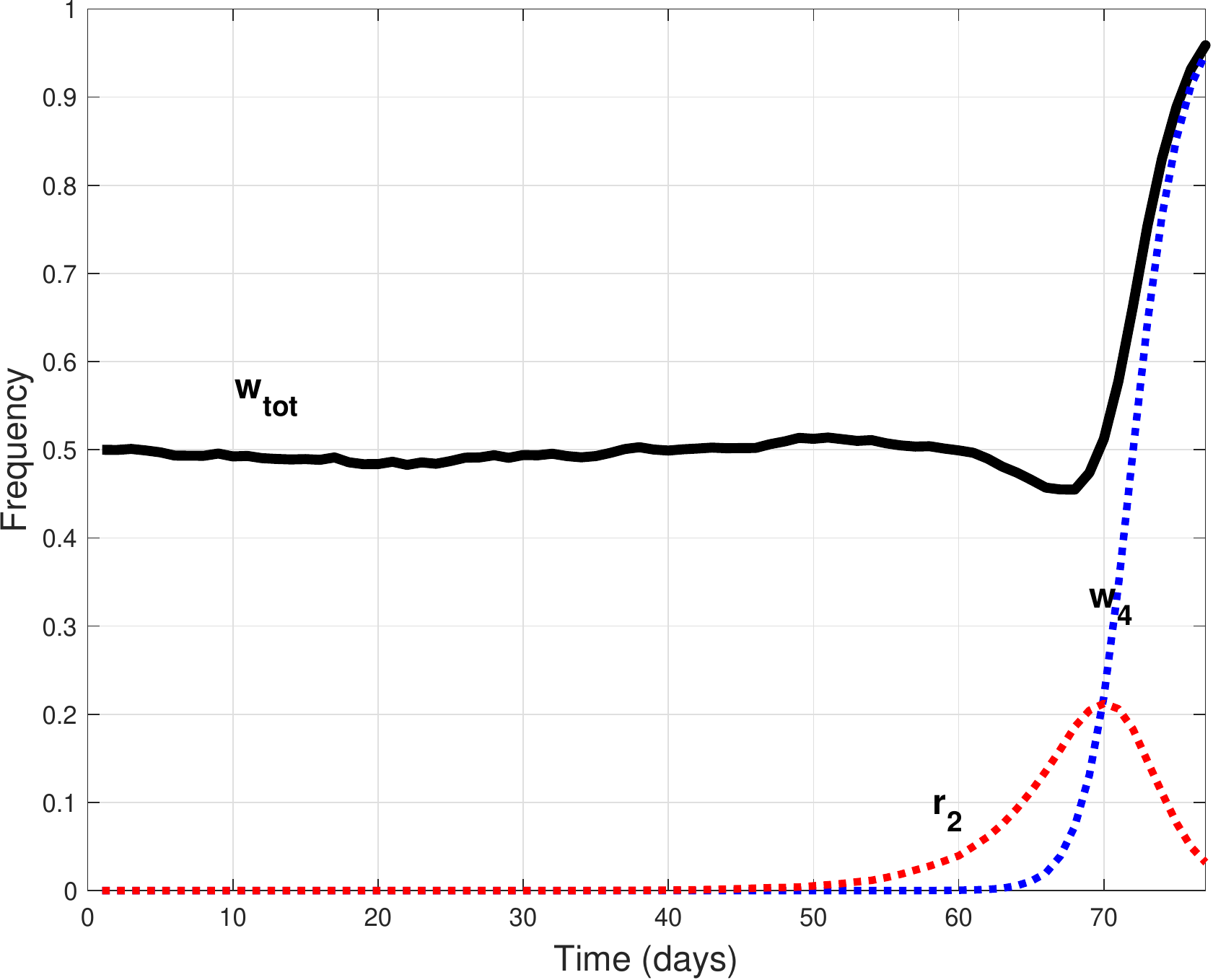}}
	\caption{Two examples of trajectories with two mutations. 
	We display the frequencies of white cells and of sub-populations of mutants.
	Black line (labeled $w_{tot}$) = total white cells frequency $w(t)$, 
	Left part - emergence of genotype-$2$ mutants among white cells (blue line $w_2$) and 
	later emergence of genotype-$3$ mutants among red cells (red line $r_3$), 
	Right part - emergence of  genotype-$2$ mutants among red cells (red line $r_2$) and 
	later emergence of genotype-$4$ mutants among white cells (blue line $w_4$).}
	\label{fig3a}
\end{figure}

Examples of trajectories with one mutation and two mutations are presented in Figures \ref{fig3} and \ref{fig3a}, respectively. We consider that a trajectory reaches  \emph{fixation} at time $T_{end}$ if either $w(T_{end}) \ge 95\%$ or $ w(T_{end}) \le 5\%$ and our Figures thus involve only the days $t \leq T_{end}$. For brevity we call \emph{mutation event} the emergence of a new mutant strain before fixation time.

We considered that a mutation of type $j$-white had actually emerged at time $\tau < T_{end}$, if $\tau$  is the smallest time $t$ such that $w_j(t) \geq 1/1000$. Same convention for mutations of type $j$-red cells. Among the $n=1000$ simulated trajectories of duration $T=200$ days 0.5\% trajectories have no mutation event, 
70\% trajectories have only one mutation event, 
23\% trajectories have two mutation events, 
and 6.5\% trajectories have three or more mutation events.
The mean fixation time for trajectories with exactly $k$ mutation events increases 
with $k$, starting from $T_{end} \approx 70$ for $k=1$ and reaches 
the value $T_{end} \approx 180$ for $k=3$.
We also simulated a much larger number of trajectories $n=100,000$ to estimate
the frequency of emergence $f(j)$ for mutant genotypes $j=2,\ldots,4$.
This yielded $f(2) = 19\%$, $f(3) = 29\%$,  $f(4) = 52\%$. This implies that 
more mutation events is recorded for stronger genotypes and, therefore, 
the error of estimating selective advantages 
would decrease for mutant genotypes with larger fitnesses (bigger $j$). 

\subsection{Selective Advantages Estimation based on 1,000 trajectories}
\label{sec:num}

To validate our estimation approach we use only data for the total frequency of white cells, $w(t)$, and we first estimate approximate time of the first mutation (see section \ref{sec:T1}) for each trajectory.
Out of $n=1000$ simulated trajectories, 995 trajectories exhibit at least one mutation event.
Thus, we apply our least-squares algorithm to these 995 trajectories 
to estimate selective advantages of mutant genotypes (section \ref{sec5.6}) 
and then combine these estimates into the histogram $H_1$ (section \ref{sec:pool}).

\begin{figure}[t]
\centerline{\includegraphics[scale = 0.5]{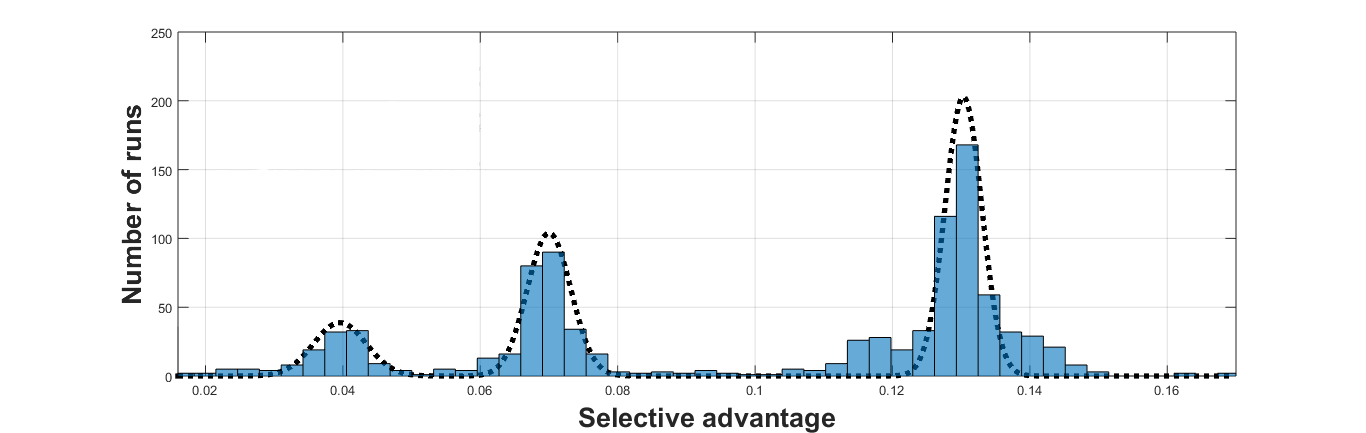}}
\caption{Histogram of the 3 selective advantages based on the first mutation event for
$995$ simulated trajectories and
corresponding fit $G(x)$ in \eqref{multiG}. Corresponding estimates for the selective advantages are $\hat{s}_2 = 0.039$, $\hat{s}_3 = 0.073$, and $\hat{s}_4 = 0.134$.}
\label{fig1}
\end{figure}

Histogram $H_1$ and the corresponding fit $G(x)$ in \eqref{multiG} are depicted in Figure \ref{fig1}.
Histogram $H_1$ has clear three local maxima with corresponding estimates for selective advantages
$\hat{s}_2 = 0.039$, $\hat{s}_3 = 0.073$, and $\hat{s}_4 = 0.134$. 
These estimation results are in a very good agreement with true parameter values $s_2 = 0.04$, $s_3 =0.07$, and $s_4= 0.13$. 
Estimation errors for all three parameters are less than 5\%.
Standard deviations for the multi-Gaussian fit 
$(c_2, c_3, c_4) \approx$ $(0.012, 0.006, 0.005)$, 
provide estimates for the relative accuracy of estimation of selective advantages.
Here we can clearly see that selective advantages of the two stronger phenotypes are estimated more
accurately than the selective advantage of the two weaker mutant phenotype.

%
%

In addition to the analysis of the first mutation event, we also analyzed the emergence of 
second mutation in 995 trajectories as discussed in sections \ref{sec:T2} and \ref{sec:twoevent}.
Approximately 300 trajectories exhibit a second mutation event and for each one of those trajectories, we applied our non-linear least squares algorithms to analyze the second emergence of the second mutant strain. 
This generated a histogram $H_2$. However, histogram $H_2$ in this case 
has only one clearly pronounced peak.
Moreover, multi-Gaussian fit for the histogram $H_2$ resulted in much larger (at least one order of magnitude) standard deviations for the standard deviations $c_j$, $j=2,\ldots,4$. Therefore, 
we concluded that $300$ trajectories are not sufficient to accurately estimate selective advantages based on the second mutation event and combine histograms $H_1$ and $H_2$.

\subsection{Selective Advantages Estimation based on 10,000 trajectories}
To evaluate more precisely the impact of the second mutation event analysis on selective advantages estimates, we also performed estimation of selective advantages from $n=10,000$ trajectories with duration 200 days each.
This resulted in approximately 3700 trajectories exhibiting  at least 2 mutation events.

\begin{figure}[t]
\centerline{\includegraphics[scale = 0.5]{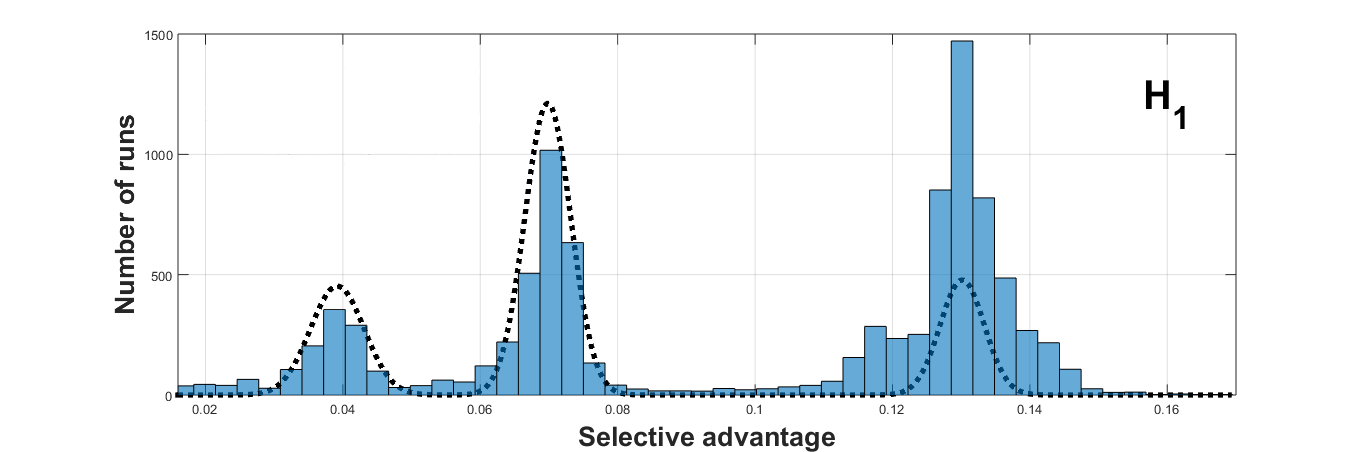}}
\centerline{\includegraphics[scale = 0.5]{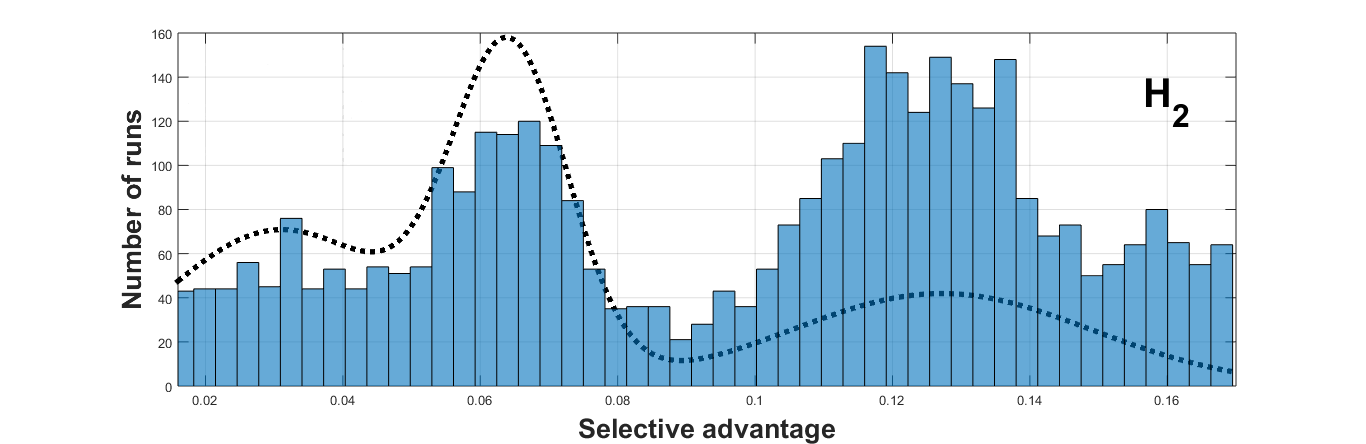}}
\centerline{\includegraphics[scale = 0.5]{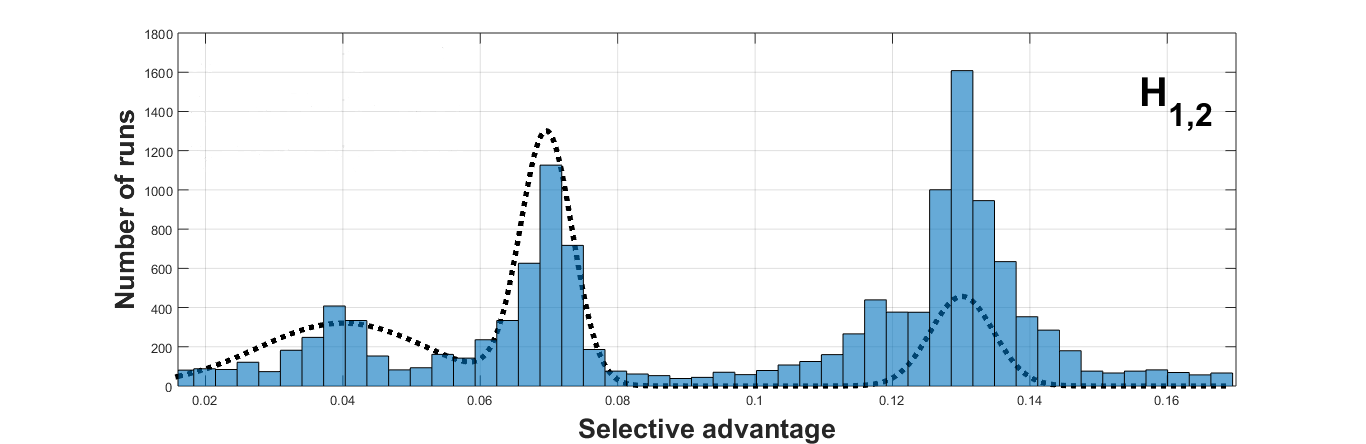}}
\caption{Gaussian mixture estimates  for selective advantages. Results are based on algorithmic analysis of  $n=10,000$ simulated trajectories.
Top : Estimates of selective advantages by analysis of first new mutant strain; 
Middle : Estimation of selective advantages by analysis of second new mutant strain; 
Bottom : Estimation of selective advantages by joint analysis of first and second new mutant strain.}
\label{fig2}
\end{figure}

In figure \ref{fig2}, histograms $H_1$ (top) and $H_2$ (middle) depict 
results of estimation based on the first and second mutation events, respectively.
Histogram $H_{1,2}$ is obtained by combining $H_1$ and $H_2$.
There are much fewer trajectories exhibiting at least two mutation events compared with 
trajectories exhibiting at least one mutation event. 
Therefore, estimates based on the second mutation event have much larger standard errors
than the $H_1$ estimates. Although the number of trajectories exhibiting a second mutation event is relatively high, the combined histogram $H_{1,2}$ provides considerably less accurate
estimates for selective advantages, especially for the genotype $j=2$.
We would like to point out that these results are particular to the mutation matrix $P$ in \eqref{P}.
In particular, for the mutation matrix $P$ considered here all mutant genotypes ($j > 1$) can emerge directly from the ancestor genotype $j=1$ with equal probabilities $1/(g-1)$. Therefore, we can conjecture 
that in such cases analysis of the first mutation event contains sufficient information to
adequately estimate selective advantages if there are
many trajectories exhibiting the first mutation event. However, there are several
situations where analysis of the second mutation event can provide useful information. We comment about this in the conclusions.

We also analyzed the behavior of standard error of estimations for selective advantages 
as the number $n$ of simulated trajectories increases. In particular, 
standard errors $c_j(n)$ with $j > 1$ were computed for $n= 75$, $150$, $250$, $1000$, $10000$, $50000$, $100000$. A linear regression demonstrated 
that for $n \geq 250$ standard errors decay as $c_j(n) \sim n^{-1/2}$. 
It is possible to provide an analytical argument which supports this conclusion for large $n$.

\section{Conclusions}
\label{sec:conclusions}
In this paper we present an efficient  non-linear least squares approach to estimate selective advantages of mutant genotypes, based on the type of observed data provided by long-term experiments focused on bacterial genetic evolution. In many such experiments, bacterial cells are colored with two biomarkers (e.g. white and red), and on each "day" $t$ the frequency of white and red cells is recorded.
We analyze a Markov chain model which corresponds to this experimental setup for evolutionary experiments. The dynamics of this model is based on daily cycles which include three phases - deterministic growth, random mutations, and random selection (dilution). 
The stochastic model involves a finite set of $g$ distinct genotypes 
with one ancestor genotype and $g-1$ mutant genotypes with unknown 
selective advantages.

Typically, evolutionary experiments involve $n$ independent populations evolving in parallel, all starting with a population of $N$ cells having the same ancestor genotype. 
Therefore, such evolutionary experiments usually generate 
an ensemble of time-series (trajectories) of observed white cells frequencies.
For each trajectory,
we perform least-squares estimation of selective advantages by minimizing the 
mean squared error between the observed white-cell frequency, $w(t)$, and prediction of the Markov chain model, $W(t)$.
We then gather these $n$ estimates of selective advantages over all observed trajectories to generate a histogram of estimated selective advantages.
Typically, we expect this histogram to have several peaks which correspond 
to selective advantages of mutant genotypes.
We then approximate the histogram by a multi-Gaussian mixture to estimate 
values of selective advantages. In addition, variances in the muti-Gaussian approximation provide rough estimate on the uncertainty in the estimation
of the corresponding selective advantages.

We demonstrated applicability of our approach by applying it to synthetic datasets generated by the Markov chain model discussed above. 
We verified that our estimation approach is able to recover the 
selective advantages used in the model. As an outcome of our numerical investigation 
we can can conclude that our approach produces accurate results for ensembles sizes of 100 or larger. Our approach can be applied to smaller ensembles, but estimation errors can become quite large.
In this paper we present numerical results with $g=4$ genotypes. We also validated our approach for $g=6$ genotypes.
In addition, we also validated that errors of estimation 
decrease at speed proportional to the square root of the ensemble size, 
and that the estimates for selective advantages are reasonably
accurate as soon as the ensemble size is bigger than 200.

Mathematical analysis of the evolutionary model presented in this paper enables 
estimation of selective advantages using the first mutation event or a combination of the first and second mutation events. Our numerical results demonstrate that 
a straightforward pooling of estimators obtained from the 
first and second mutation events may not be always optimal.
Instead, analysis of the second mutation event can be used to extend results
of the estimation from the first mutation event data.
However, analysis of the second mutation event can be quite useful for improving 
(or even discovering) new mutant types and/or improving 
a subset of the selective advantages estimates. First, 
second mutation events can be used to identify genotypes which were rarely seen in first mutations events, 
but more frequently present among second mutation events.
Such situations can occur 
for small ensemble sixes $n \approx 100,\ldots, 200$.
Second, 
analysis of the second mutation event
can shed light on the shape of the mutation matrix $P$ and whether all mutant genotypes 
can emerge directly from the ancestor genotype. In other words, analysis 
of the second mutation event can potentially elucidate importance of double mutations. 
Finally we have also successfully tested that our approach can easily be extended to the estimation of the mutation rate $\mu$. Additional simulations and parameter estimation studies can be found in \cite{Sarkisov2017}.

Realistic evolutionary experiments include a rather small number of populations growing in parallel. However, these experiments are carried out for many years, and bacterial populations are often re-seeded with the original ancestor genotypes after a dominant mutation occurs and one color overtakes the whole population. Therefore, a realistic number of trajectories available for estimating selective advantages are of the order of 100. Nevertheless, 
estimation procedure described here is still applicable even with such small number 
of available trajectories. 

For small number of trajectories 
our approach should then be applied sequentially as follows.
First, compute and analyze histogram $H_1$ derived from first mutation 
events and compute first estimates of selective advantages as well 
as their standard estimation errors. Second, repeat optimization procedure for 
computing estimates for selective advantages based on first mutation events  (section \ref{sec5.6})
with improved search intervals for $s \in RSEL$. In particular, $RSEL$ can be taken
as union of confidence intervals for selective advantages computed on the first pass using
multi-Gaussian fit of $H_1$. We expect that the second pass of the optimization procedure
based on the first mutation event data should yield much more accurate estimates for
time of the first mutation, $\tau_1$, selective advantage $s_{g_1}$, and the number of emerged mutants, $\gamma_1$. In addition, second pass should also yield an improved histogram
of selected advantages, $H_1$. Third, perform optimization procedure for the second mutation event data with $s \in RSEL$, where $RSEL$ has been computed from the improved $H_1$.
In contrast with the present paper where 
analysis of the second mutation event is based on the selective advantage 
of the first mutant genotype computed from a current (single) trajectory, 
using estimates from $H_1$ is more robust and should provide 
a more accurate version of $H_2$, with relatively small standard errors. 
The least squares fitting of the multi-Gaussian mixture for $H_2$ should then be performed conditional on the information gathered by the $H_1$ estimates and their standard errors. This approach will provide better estimates of selective advantages derived from $H_2$. 
The fusion of estimates obtained from $H_1$  
and $H_2$ should be performed on the basis of their respective standard errors.
We plan to implement and test this upgraded approach in a subsequent paper, applying our estimation procedure to realistic observational data.

\section*{Declaration of Interest}

The authors declare that they have no known competing financial interests or personal relationships that could have appeared to influence the work reported in this paper.

\section*{Acknowledgment}

This research has been partially supported by NSF grants DMS-1412927 and DMS-1903270.

\clearpage

%

\end{document}